\pdfoutput=1

\RequirePackage{fix-cm}

\documentclass[twocolumn]{svjour3}

\smartqed

\usepackage{graphicx,amssymb}
\usepackage{epsf,psfig}
\usepackage{txfonts}
\usepackage{natbib}

\journalname{Nonlinear Dynamics}

\begin{document}

\title{A Hamiltonian system of three degrees of freedom with eight channels of escape: The Great Escape}

\author{Euaggelos E. Zotos}

\institute{Department of Physics, School of Science, \\
Aristotle University of Thessaloniki, \\
GR-541 24, Thessaloniki, Greece \\
Corresponding author's email: {evzotos@physics.auth.gr}
}

\date{Received: 6 November 2013 / Accepted: 17 December 2013 / Published online: 10 January 2014}

\titlerunning{A Hamiltonian system of three degrees of freedom with eight channels of escape: The Great Escape}

\authorrunning{E. E. Zotos}

\maketitle

\begin{abstract}

In this work, we try to shed some light to the nature of orbits in a three-dimensional potential of a perturbed harmonic oscillator with eight possible channels of escape, which was chosen as an interesting example of open three-dimensional Hamiltonian systems. In particular, we conduct a thorough numerical investigation distinguishing between regular and chaotic orbits as well as between trapped and escaping orbits, considering unbounded motion for several values of the energy. In an attempt to discriminate safely and with certainty between ordered and chaotic motion, we use the Smaller ALingment Index (SALI) detector, computed by integrating numerically the basic equations of motion as well as the variational equations. Of particular interest, is to locate the basins of escape towards the different escape channels and connect them with the corresponding escape periods of the orbits. We split our study into three different cases depending on the initial value of the $z$ coordinate which was used for launching the test particles. We found, that when the orbits are started very close to the primary $(x,y)$ plane the respective grids exhibit a high degree of fractalization, while on the other hand for orbits with relatively high values of $z_0$ several well-formed basins of escape emerge thus, reducing significantly the fractalization of the grids. It was also observed, that for values of energy very close to the escape energy the escape times of orbits are large, while for energy levels much higher than the escape energy the vast majority of orbits escape extremely fast or even immediately to infinity. We hope our outcomes to be useful for a further understanding of the escape process in open 3D Hamiltonian systems.


\end{abstract}

\section{Introduction}
\label{intro}

The issue of escapes in Hamiltonian systems is directly related to the problem of chaotic scattering which has been an active field of research over the last decades and it still remains open (e.g., [\citealp{BTS96} -- \citealp{BGOB88}, \citealp{CPR75}, \citealp{C90}, \citealp{CK92}, \citealp{E88}, \citealp{JS88}, \citealp{ML02} -- \citealp{PH86}, \citealp{SASL06} -- \citealp{SS10}]). We know, that particular types of Hamiltonian systems have a finite energy of escape and for energy levels beyond the escape energy, these systems allow particles to escape to infinity. The literature is replete with studies of such ``open" Hamiltonian systems (e.g., [\citealp{BBS09}, \citealp{CKK93}, \citealp{KSCD99}, \citealp{SCK95} -- \citealp{SKCD96}]).

Usually, the infinity acts as an attractor for an escape particle, which may escape through different channels (exits) on the equipotential curve or on the equipotential surface depending whether the dynamical system is two or three-dimensional, respectively. Therefore, it is quite possible to obtain basins of escape, similar to basins of attraction in dissipative systems or even the Newton-Raphson fractal structures. Basins of escape have been studied in several papers (e.g., [\citealp{BGOB88}, \citealp{C02}, \citealp{KY91}, \citealp{PCOG96}]). The reader can find more details regarding basins of escape in [\citealp{C02}].

The well-known H\'{e}non-Heiles system [\citealp{HH64}] is undoubtedly a paradigmatic model for time-independent Hamiltonian systems of two degrees of freedom. A huge load of research on escape from this system has been conducted over the years (e.g., [\citealp{AJ03} -- \citealp{AVS03}, \citealp{BBS08}, \citealp{BSBS12}, \citealp{dML99}, \citealp{S07}]). Here, we would like to point out that all the above references on escapes in the H\'{e}non-Heiles system are exemplary rather than exhaustive, taking into account that a vast quantity of related literature exists.

A simple dynamical system of two coupled harmonic oscillators for various values of the energy above the escape energy has been investigated in [\citealp{CE04}], where it was found that stable periodic orbits are surrounded by stability islands that never escape. A further numerical analysis of the same dynamical system in [\citealp{CHLG12}], revealed that as the energy increases beyond the escape energy, the majority of chaotic orbits escape either directly, or after a small or large number of intersections with the $y = 0$ axis. In the same vein, the effects of different types of perturbations on both the topology and the escaping dynamics in the H\'{e}non-Heiles system was studied in [\citealp{BSBS12}], where basins of escape were found to exits in the physical $(x,y)$ as well as in the phase $(y,\dot{y})$ space.

Of particular interest is the issue of escaping orbits in galactic dynamics. In a recent article [\citealp{Z12b}], we explored the nature of the orbits of stars in a galactic-type potential, which can be considered to describe local motion in the meridional plane $(R,z)$ near the central parts of an axially symmetric galaxy. It was observed, that apart from the trapped orbits there are two types of escaping orbits, those which escape fast and those which need to spend vast time intervals inside the limiting curve before they find the exit and eventually escape. Furthermore, the chaotic dynamics within a star cluster embedded in the tidal field of a galaxy was examined in [\citealp{EJSP08}]. In particular, by scanning thoroughly the phase plane and obtaining the basins of escape with the respective escape times it was revealed, that the higher escape times correspond to initial conditions of orbits near the fractal basin boundaries.

We should like to point out, that the vast majority of the existed literature deals with escaping orbits in Hamiltonian systems of two degrees of freedom, while only a handful of papers is devoted to escapes in dynamical systems with three degrees of freedom. Thus, we decided to make a new contribution by exploring the nature of orbits and the escape process in a simple three-dimensional (3D) potential with eight possible channels of escape. The aim of this work, is twofold: (i) to distinguish between ordered/chaotic and trapped/escaping orbits and (ii) to locate the basins of escape leading to different escape channels and try to connect them with the corresponding escape times of the orbits.

Over the last half century, dynamical systems made up of perturbed harmonic oscillators have been extensively used in order to describe local motion (i.e., near an equilibrium point) (e.g., [\citealp{AEFR06}, \citealp{C93}, \citealp{CK98}, \citealp{CK99}, \citealp{FLP98a} -- \citealp{HH64}, \citealp{SI79}, \citealp{Z12c}]). In an attempt to reveal and understand the nature of orbits in these systems, scientists have used either numerical (e.g., [\citealp{KV08}, \citealp{ZC12}]) or analytical methods (e.g., [\citealp{CB82}, \citealp{D91}, \citealp{DE91}, \citealp{E00}, \citealp{ED99}]). Furthermore, potentials made up of harmonic oscillators are frequently used in Astronomy, as a first step for distinguishing between ordered and chaotic local motion in galaxies, since it is widely accepted that the motion of stars near the central region of a galaxy can be approximated by harmonic oscillations. Therefore, the above-mentioned facts justify our choice of using a potential of a perturbed harmonic oscillator in our quest for escaping orbits.

The structure of the present article is as follows: in Section \ref{modpot} we describe the properties of the potential we chose for our investigation of escaping orbits. The computational methods used in order to determine the nature (ordered/chaotic and trapped/escaping) of orbits are described in Section \ref{cometh}. In the following Section, we conduct a thorough analysis of the orbits presenting in detail all the numerical results of our computations. Our article ends with Section \ref{disc}, where the discussion and the conclusions of this research are presented.

\section{Properties of the model potential}
\label{modpot}

The general form of a three-dimensional perturbed harmonic oscillators is
\begin{equation}
V(x,y,z) = \frac{1}{2}\left(\omega_1^2 x^2 + \omega_2^2 y^2 + \omega_3^2 z^2 \right) + \varepsilon V_1(x,y,z),
\label{genform}
\end{equation}
where $\omega_1$, $\omega_2$ and $\omega_3$ are the unperturbed frequencies of oscillations along the $x$, $y$ and $z$ axes respectively, $\varepsilon$ is the perturbation parameter, while $V_1$ is the function containing the perturbing terms. This is called a three-dimensional perturbed elliptic oscillator.

In the present paper, we shall use a three-dimensional perturbed harmonic oscillator at the 1:1:1 resonance, that is when $\omega_1 = \omega_2 = \omega_3 = \omega$, in order to investigate the escape properties of orbits. The corresponding potential is
\begin{equation}
V(x,y,z) = \frac{\omega^2}{2}\left(x^2 + y^2 + z^2\right) + \varepsilon V_1(x,y,z),
\label{pot}
\end{equation}
where in our case the perturbation is given by
\begin{equation}
V_1(x,y,z) = x^2y^2 + y^2z^2 + x^2z^2 - x^2y^2z^2,
\label{pert}
\end{equation}
being $\omega$ the common frequency of oscillations along the three axes. Without the loss of generality, we may set $\omega = 1$ and $\varepsilon = 1$ for more convenient numerical computations. The same potential was also used in [\citealp{CZ12}] however, the study of the nature of orbits was restricted only to bounded motion.

\begin{figure*}
\centering
\resizebox{0.7\hsize}{!}{\includegraphics{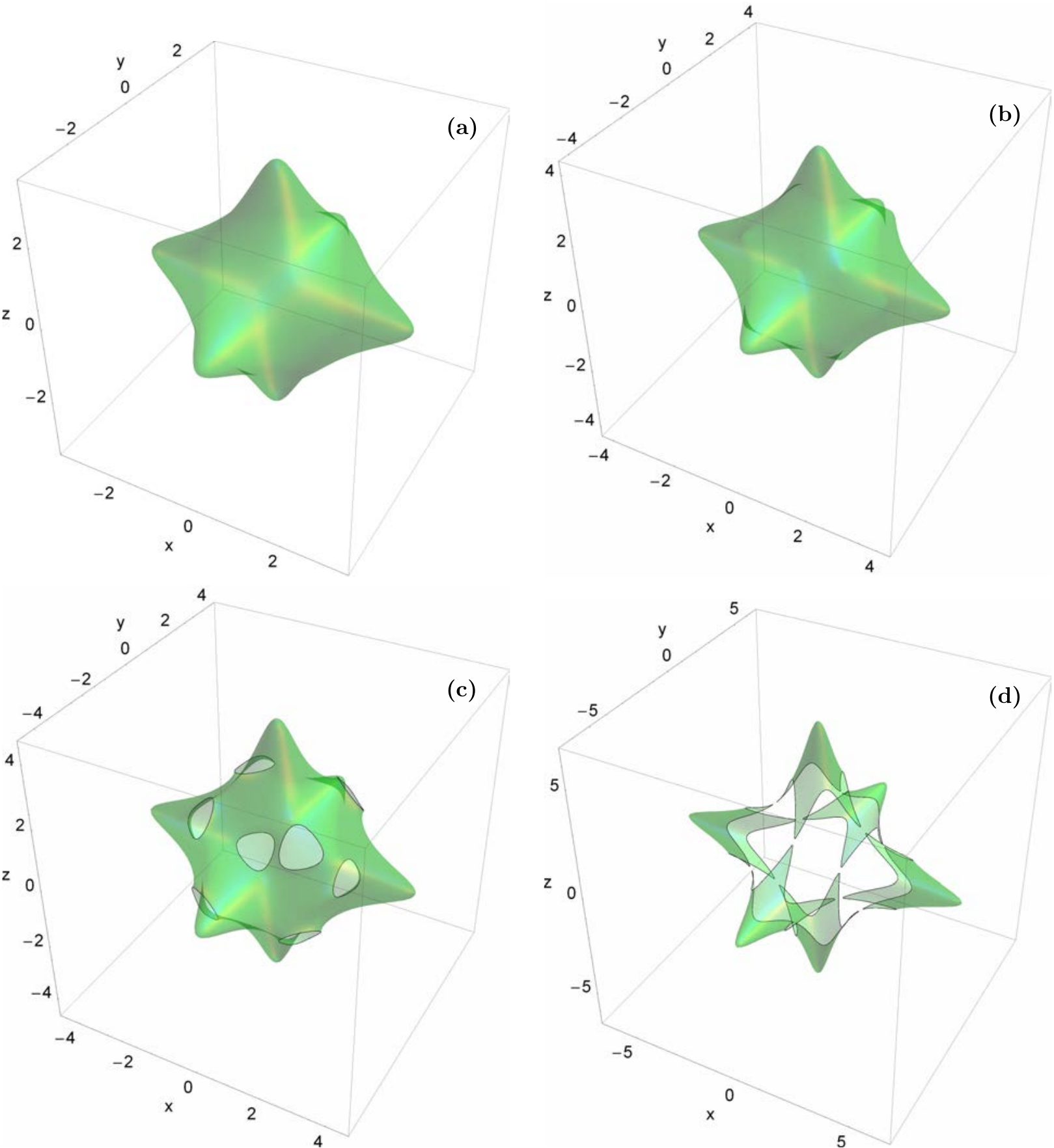}}
\caption{Plot of the isopotential surface $V(x,y,z) = h$, when (a-upper left): $h = 5 < h_{\rm esc}$, (b-upper right): $h = h_{\rm esc}$, (c-lower left): $h = 9 > h_{\rm esc}$ and (d-lower right): $h = 20 \gg h_{\rm esc}$.}
\label{surfs}
\end{figure*}

The Hamiltonian to potential (\ref{pot}) reads
\begin{equation}
H = \frac{1}{2}\left(\dot{x}^2 + \dot{y}^2 + \dot{z}^2 \right) + V(x,y,z) = h,
\label{ham}
\end{equation}
where $\dot{x}$, $\dot{y}$ and $\dot{z}$ are the momenta per unit mass conjugate to $x$, $y$ and $z$ respectively, while $h > 0$ is the numerical value of the Hamiltonian, which is conserved.

The basic equations of motion for a test particle with a unit mass are
\begin{equation}
\ddot{x} = - \frac{\partial V}{\partial x}, \ \ \
\ddot{y} = - \frac{\partial V}{\partial y}, \ \ \
\ddot{z} = - \frac{\partial V}{\partial z},
\label{eqmot}
\end{equation}
where, as usual, the dot indicates derivative with respect to the time. Furthermore, the variational equations governing the evolution of a deviation vector $\vec{w} = (\delta x, \delta y, \delta z, \delta \dot{x}, \delta \dot{y}, \delta \dot{z})$ are
\begin{eqnarray}
\dot{(\delta x)} &=& \delta \dot{x}, \ \ \
\dot{(\delta y)} = \delta \dot{y}, \ \ \
\dot{(\delta z)} = \delta \dot{z}, \nonumber \\
(\dot{\delta \dot{x}}) &=& -\frac{\partial^2 V}{\partial x^2}\delta x - \frac{\partial^2 V}{\partial x \partial y}\delta y -\frac{\partial^2 V}{\partial x \partial z}\delta z, \nonumber \\
(\dot{\delta \dot{y}}) &=& -\frac{\partial^2 V}{\partial y \partial x}\delta x - \frac{\partial^2 V}{\partial y^2}\delta y -\frac{\partial^2 V}{\partial y \partial z}\delta z, \nonumber \\
(\dot{\delta \dot{z}}) &=& -\frac{\partial^2 V}{\partial z \partial x}\delta x - \frac{\partial^2 V}{\partial z \partial y}\delta y -\frac{\partial^2 V}{\partial z^2}\delta z. \nonumber \\
\label{variac}
\end{eqnarray}

\begin{figure}
\includegraphics[width=\hsize]{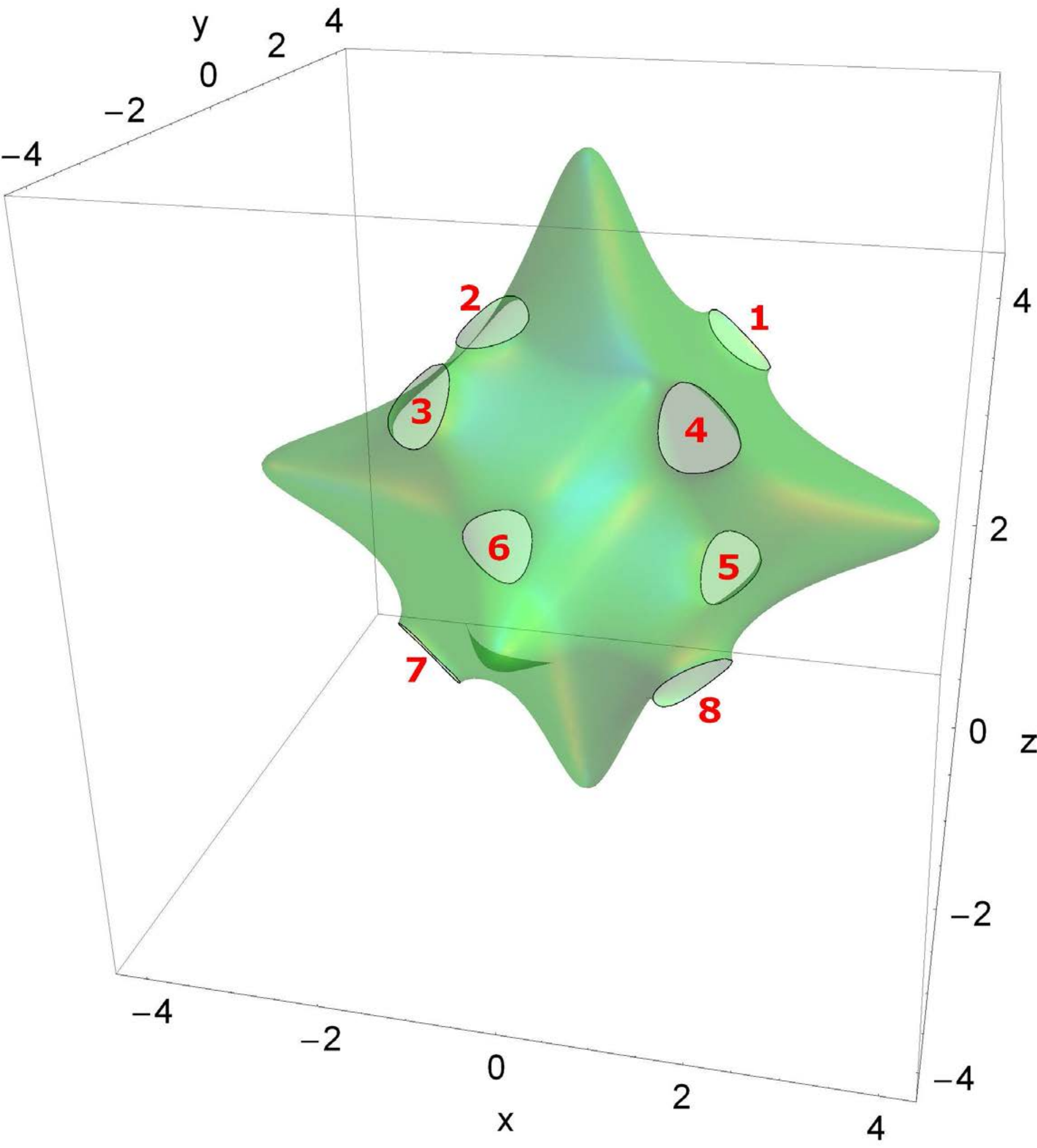}
\caption{The eight symmetrical channels of escape when the motion is unbounded $(h > h_{\rm esc})$.}
\label{chans}
\end{figure}

\begin{figure}
\includegraphics[width=\hsize]{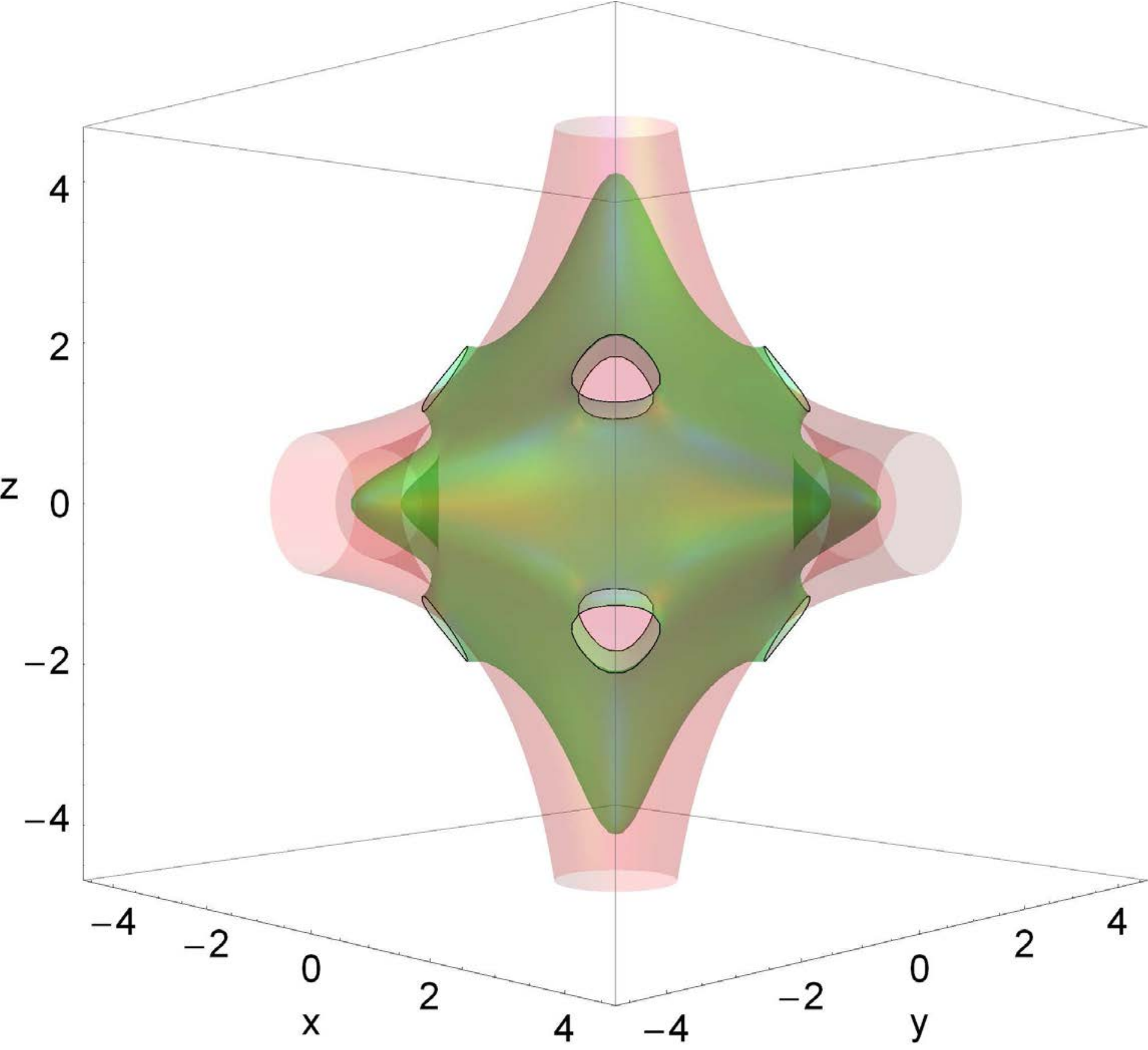}
\caption{A plot showing together the isopotential and the cutoff surface which determines the escape of orbits. An orbit is considered as escaping when it crosses one of the eight channels and intersects the cutoff surface, with velocity pointing outwards.}
\label{cutoff}
\end{figure}

\begin{figure}
\includegraphics[width=\hsize]{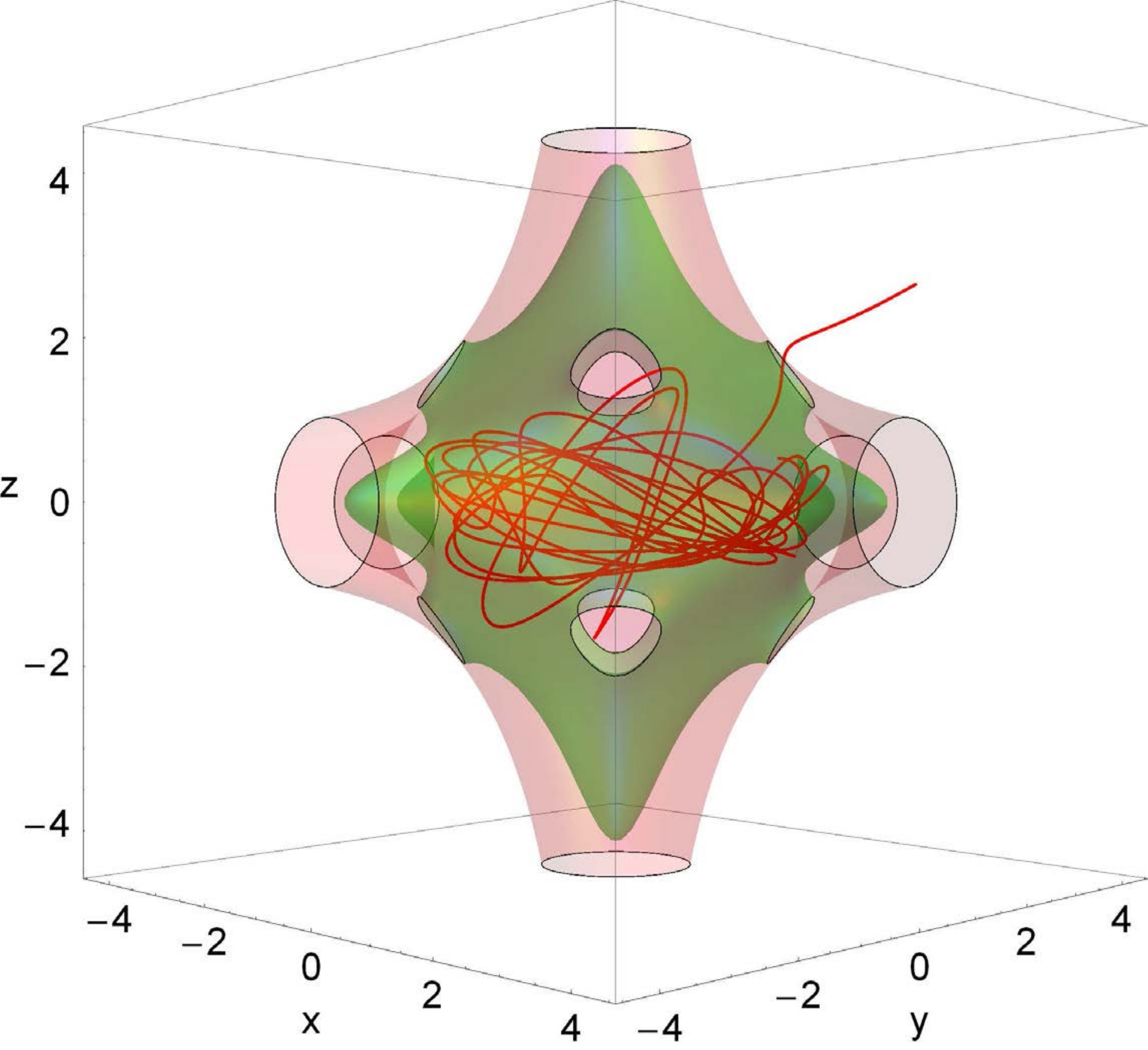}
\caption{A characteristic example of a three-dimensional orbits which escapes to infinity from escape channel 1.}
\label{orbesc}
\end{figure}

Potential (\ref{pot}) has a finite energy of escape which is given by
\begin{equation}
h_{\rm esc} = \frac{\omega^2}{2} + c \left(1 + \sqrt{\frac{c}{2\varepsilon}} \right),
\label{hesc}
\end{equation}
where $c = \omega^2 + 2\varepsilon$. Eq. (\ref{hesc}) is a general formula, while in our case where $\omega = \varepsilon = 1$ the escape energy takes the value $h_{\rm esc} = 7.1742346141747675$. For values of energy smaller than the energy of escape, the isopotential surface $V(x,y,z) = h$ is closed. We see in Fig. \ref{surfs}a where $h = 5 < h_{\rm esc}$, that the three-dimensional isopotential surface is closed and has a 3D star-like shape. In Fig. \ref{surfs}b where $h = h_{\rm esc}$ we observe the presence of some very sharp edges indicating that the surface is about to open. Indeed, when $h = 9 > h_{\rm esc}$ it can be seen in Fig. \ref{surfs}c, that the surface opens and eight holes appear. These holes are in fact escaping channels through which a test particle is free to escape to infinity only when the motion is unbounded $(h > h_{\rm esc})$. The escaping channels are symmetrical to the three primary planes $(x = 0, y = 0, z = 0)$, due to the overall symmetry of the surface. In our investigation, we will perform a thorough statistical analysis regarding the escaping channels therefore, in Fig. \ref{chans} we number each channel. Even though the escape channels are symmetrical, we can distinguish between them from their particular position on the surface and the corresponding sings of the coordinates $(x,y,z)$. In particular we have: $C_1 \rightarrow (+,+,+)$, $C_2 \rightarrow (-,+,+)$, $C_3 \rightarrow (-,-,+)$, $C_4 \rightarrow (+,-,+)$, $C_5 \rightarrow (+,+,-)$, $C_6 \rightarrow (-,+,-)$, $C_7 \rightarrow (-,-,-)$ and $C_8 \rightarrow (+,-,-)$, where $+$ indicates positive coordinate, while $-$ stands for negative coordinates.

An issue of paramount importance is the determination of the position as well as the time at which an orbit escapes. In two-dimensional systems, the highly unstable Lyapunov periodic orbits control the properties of escaping orbits as they bridge the openings of the zero velocity curve. In particular, any orbit that crosses a Lyapunov orbit with velocity pointing outwards escapes from the system (see e.g., [\citealp{C90}]). As far as we know, in three-dimensional systems this method cannot be applied so, we have to find another, possibly new, way of determining the escapes of orbits. Taking into account the particular geometry of the isopotential surface, we could define an appropriate cutoff surface which embraces the isopotential and abuts on the eight escape channels (see Fig. \ref{cutoff}). After conducting extensive numerical simulations we found, that the best cutoff surface is the following
\begin{equation}
f_{\rm c}(x,y,z) = \sqrt{x^2y^2 + y^2z^2 + x^2z^2}.
\label{fcut}
\end{equation}
Thus, a three-dimensional orbit is considered to escape one when $f_{\rm c} \geq \sqrt{h}$. In Fig. \ref{orbesc} we present a characteristic example of an orbit which escapes from channel 1.

In our investigation, we shall deal only with unbounded motion of test particles for values of energy in the set $h = \{7.5,8.5,9.5, ..., 18.5\}$. Our numerical calculations indicate, that when the value of the energy is large enough $h > 19.2$ the openings are so wide, that as we can see in Fig. \ref{surfs}d where $h = 20 \gg h_{\rm esc}$ the isopotential surface is dismembered into six distinct components and therefore, it is no longer possible to define escapes through particular channels.

\section{Computational methods}
\label{cometh}

\begin{figure*}
\centering
\resizebox{0.8\hsize}{!}{\includegraphics{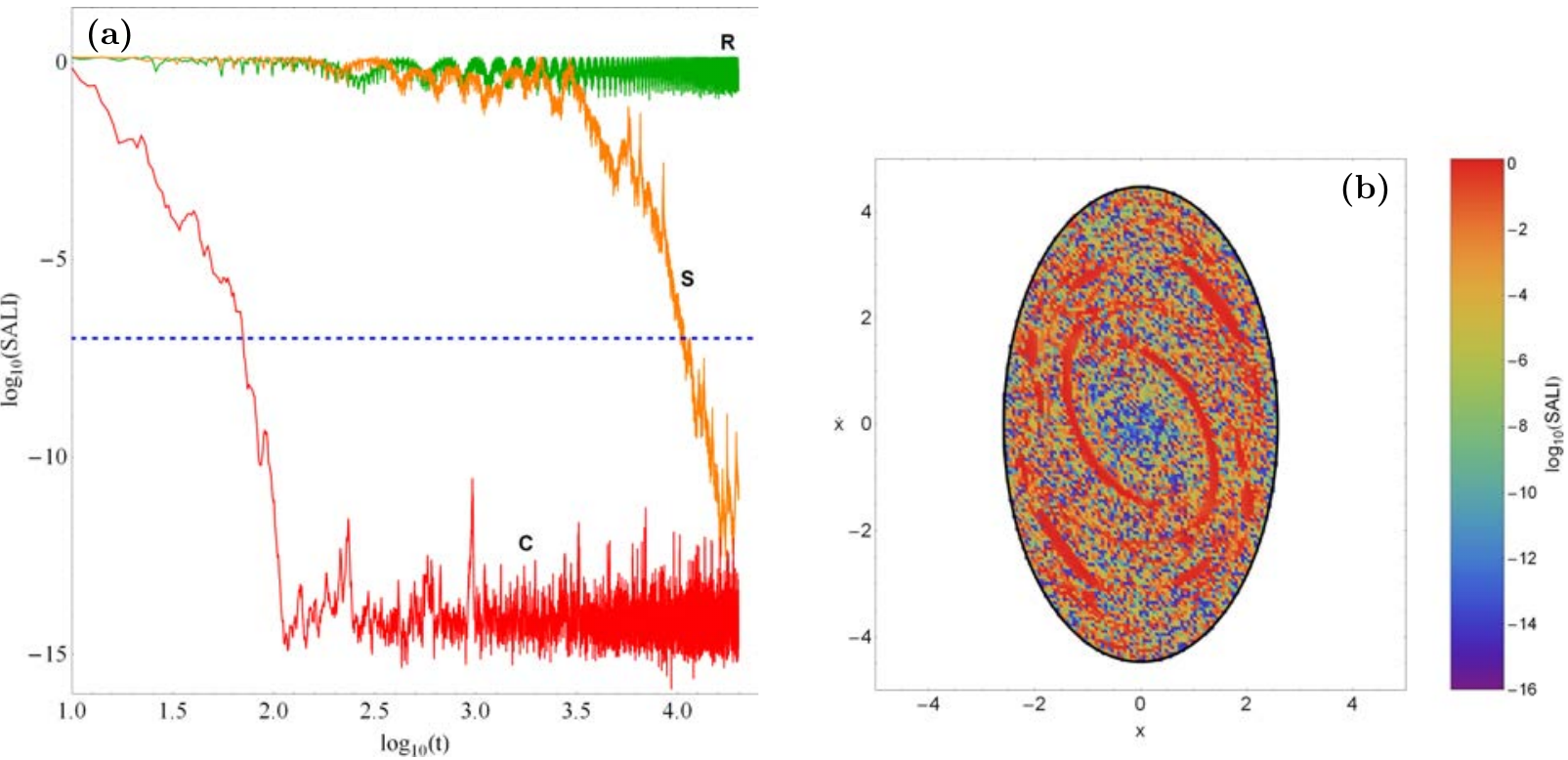}}
\caption{(a-left): Time-evolution of the SALI of a regular orbit (green color - R), a sticky orbit (orange color - S) and a chaotic orbit (red color - C) in our model for a time period of $2 \times 10^4$ time units. The horizontal, blue, dashed line corresponds to the threshold value $10^{-7}$ which separates regular from chaotic motion. The chaotic orbit needs only about 70 time units in order to cross the threshold, while on the other hand, the sticky orbit requires a considerable longer integration time of about 10000 time units so as to reveal its true chaotic nature. (b-right): Regions of different values of the SALI on the $(x,\dot{x})$ plane. Light reddish colors correspond to ordered motion, while dark blue/purplr colors indicate chaotic orbits.}
\label{salis}
\end{figure*}

Knowing the ordered or chaotic nature of orbits (trapped and escaping) is an issue of great interest. Over the years, several chaos indicators have been developed in order to determine the character of orbits. In our case, we chose to use the Smaller ALingment index (SALI) method. The SALI [\citealp{S01}, \citealp{SABV04}] has been proved a very fast, reliable and effective tool, which is defined as
\begin{equation}
\rm SALI(t) \equiv min(d_-, d_+),
\label{sali}
\end{equation}
where $d_- \equiv \| {\bf{w_1}}(t) - {\bf{w_2}}(t) \|$ and $d_+ \equiv \| {\bf{w_1}}(t) + {\bf{w_2}}(t) \|$ are the alignments indices, while ${\bf{w_1}}(t)$ and ${\bf{w_2}}(t)$, are two deviations vectors which initially point in two random directions. For distinguishing between ordered and chaotic motion, all we have to do is to compute the SALI for a relatively short time interval of numerical integration $t_{max}$. In particular, we track simultaneously the time-evolution of the main orbit itself as well as the two deviation vector ${\bf{w_1}}(t)$ and ${\bf{w_2}}(t)$ in order to compute the SALI. The variational equations (\ref{variac}), as usual, are used for the evolution and computation of the deviation vectors.

The time-evolution of SALI strongly depends on the nature of the computed orbit since when the orbit is regular the SALI exhibits small fluctuations around non zero values, while on the other hand, in the case of chaotic orbits the SALI after a small transient period it tends exponentially to zero approaching the limit of the accuracy of the computer $(10^{-16})$. Therefore, the particular time-evolution of the SALI allow us to distinguish fast and safely between regular and chaotic motion. The time-evolution of a regular (R) and a chaotic (C) orbit for a time period of $2 \times 10^4$ time units is presented in Fig. \ref{salis}a. We observe, that both regular and chaotic orbits exhibit the expected behavior. Nevertheless, we have to define a specific numerical threshold value for determining the transition from regularity to chaos. After conducting extensive numerical experiments, integrating many sets of orbits, we conclude that a safe threshold value for the SALI is the value $10^{-7}$. The horizontal, blue, dashed line in Fig. \ref{salis}a corresponds to that threshold value which separates regular from chaotic motion. In order to decide whether an orbit is regular or chaotic, one may use the usual method according to which we check after a certain and predefined time interval of numerical integration, if the value of SALI has become less than the established threshold value. Therefore, if SALI $\leq 10^{-7}$ the orbit is chaotic, while if SALI $ > 10^{-7}$ the orbit is regular. In Fig. \ref{salis}b we present a dense grid of initial conditions $(x_0,\dot{x_0})$, where the initial conditions are colored according to the value of SALI. We clearly observe several regions of regularity indicated by light reddish colors as well as a unified chaotic domain (blue/purple dots). Therefore, the distinction between regular and chaotic motion is clear and beyond any doubt when using the SALI method.

A simple qualitative way for distinguishing between regular and chaotic motion in a Hamiltonian system is by plotting the successive intersections of the orbits using a Poincar\'{e} Surface of Section (PSS) [\citealp{HH64}]. This method has been extensively applied to 2D models, as in these systems the PSS is a two-dimensional plane. In 3D systems, however, the PSS is four-dimensional and thus the behavior of the orbits cannot be easily visualized. One way to overcome this issue is to project the PSS to phase spaces with lower dimensions, following the method used in [\citealp{Z12a}, \citealp{Z12c}, \citealp{ZC13}]. Let us start with initial conditions on a 4D grid of the PSS. In this way, we are able to identify again regions of order and chaos, which may be visualized, if we restrict our investigation to a subspace of the whole 6D phase space. We consider orbits with initial conditions $(x_0, z_0, \dot{x_0})$, $y_0 = \dot{z_0} = 0$, while the initial value of $\dot{y_0}$ is always obtained from the energy integral (\ref{ham}). In particular, we define a value of $z_0$, which is kept constant and then we calculate the SALI of the 3D orbits with initial conditions $(x_0, \dot{x_0})$, $y_0 = \dot{z_0} = 0$. Thus, we are able to construct again a 2D plot depicting the $(x, \dot{x})$ plane but with an additional value of $z_0$, since we deal with 3D motion. All the initial conditions of the 3D orbits lie inside the limiting curve defined by
\begin{equation}
f(x,\dot{x};z_0) = \frac{1}{2}\dot{x}^2 + V(x, y=0, z=z_0) = h.
\label{zvc}
\end{equation}

For the study of our models, we need to define the sample of orbits whose properties (chaos or regularity, escaping or trapped) we will identify. The best method for this purpose, would have been to choose the sets of initial conditions of the orbits from a distribution function of the model. This, however, is not available so, we define, for each set of values of the energy, a dense grid of initial conditions $(x_0, \dot{x_0})$ regularly distributed in the area allowed by the value of the energy, following the above-mentioned method. In each grid the step separation of the initial conditions along the $x$ and $\dot{x}$ axis was controlled in such a way that always there are at least 25000 orbits. For each initial condition, we integrated the equations of motion (\ref{eqmot}) as well as the variational equations (\ref{variac}) using a double precision Bulirsch-Stoer FORTRAN algorithm (e.g., [\citealp{PTVF92}]) with a small time step of order of $10^{-2}$, which is sufficient enough for the desired accuracy of our computations (i.e. our results practically do not change by halving the time step). In all cases, the energy integral (Eq. (\ref{ham})) was conserved better than one part in $10^{-10}$, although for most orbits it was better than one part in $10^{-11}$.

In our computations, we set $10^5$ time units as a maximum time of numerical integration. The vast majority of orbits (regular and chaotic) however, need considerable less time to find one of the exits in the limiting surface and eventually escape from the system (obviously, the numerical integration is effectively ended when an orbit passes through one of the escape channels and intersects the cutoff surface). Nevertheless, we decided to use such a vast integration time just to be sure that all orbits have enough time in order to escape. Remember, that there are the so called ``sticky orbits" which behave as regular ones during long periods of time. A characteristic example of a sticky orbit (S) in our dynamical model can be seen in Fig. \ref{salis}a, where we observe that the chaotic character of the particular sticky orbit is revealed only after a considerable long integration time of about 10000 time units. Here we should clarify, that orbits which do not escape after a numerical integration of $10^5$ time units are considered as non-escaping or trapped.

\section{Numerical results}
\label{numres}

Our main objective is to determine which orbits escape and which remain trapped, distinguishing simultaneously between regular and chaotic motion. Moreover, two additional properties of the orbits will be examined: (i) the channels through which the particles escape and (ii) the time-scale of the escapes (we shall also use the term escape period). In the present paper, we explore these aspects for various values of the energy $h$, as well as for the initial value of the $z$ coordinate. In particular, three different cases are considered: (a) orbits starting very close to the $(x,y)$ plane (low value of $z_0$), (b) orbits with a mediocre value of $z_0$ and (c) orbits launched at relatively large distance from the $(x,y)$ plane (high value of $z_0$).

\subsection{Case I: Orbits starting with a low value of $z_0$}
\label{case1}

\begin{figure*}
\centering
\resizebox{0.8\hsize}{!}{\includegraphics{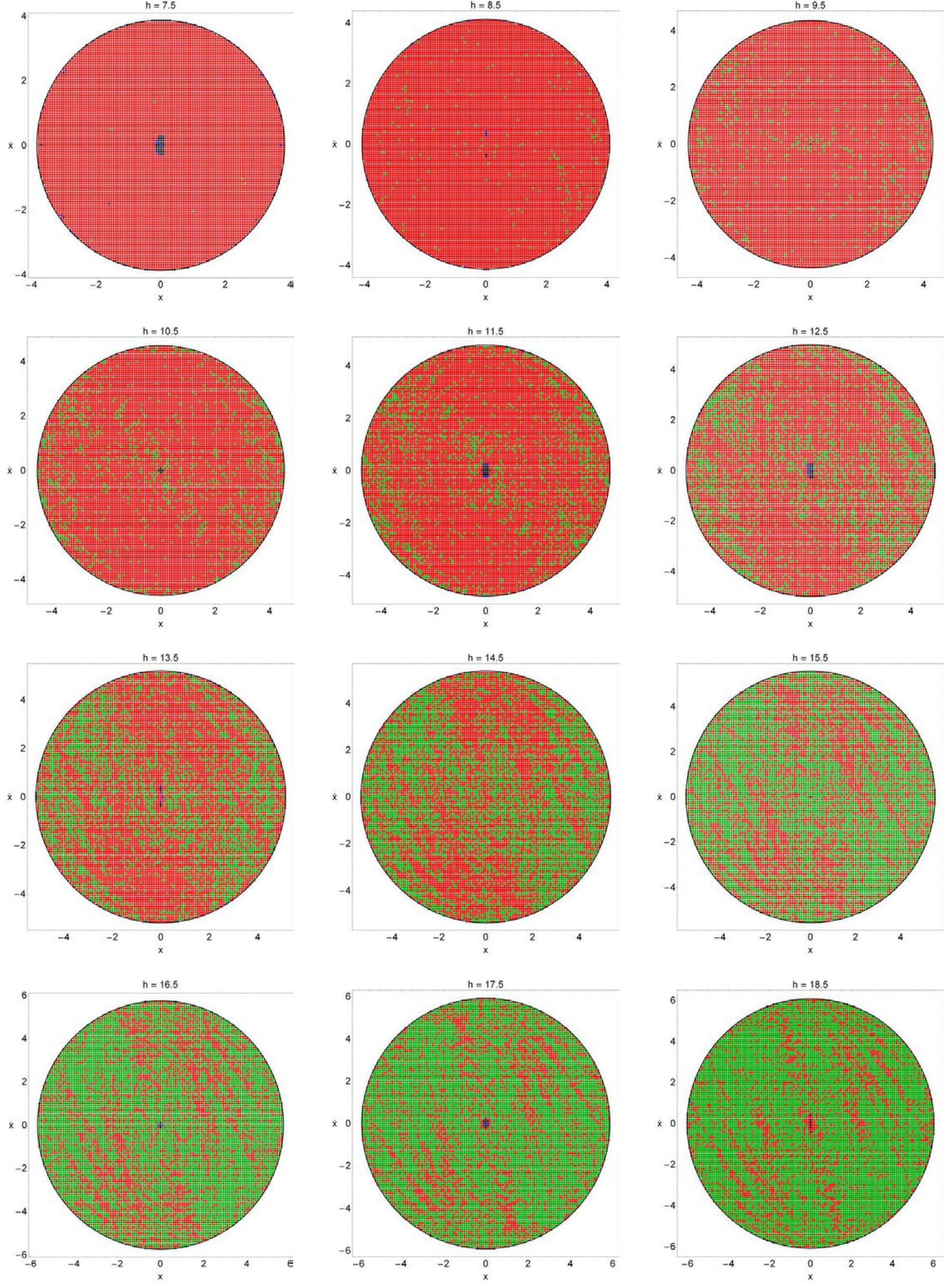}}
\caption{The structure of the $(x,\dot{x})$ plane for several values of the energy $h$, distinguishing between regular trapped orbits (black), chaotic trapped orbits (blue), regular escaping (green) and chaotic escaping (red), when $z_0 = 0.1$.}
\label{rcz01}
\end{figure*}

\begin{figure}
\includegraphics[width=\hsize]{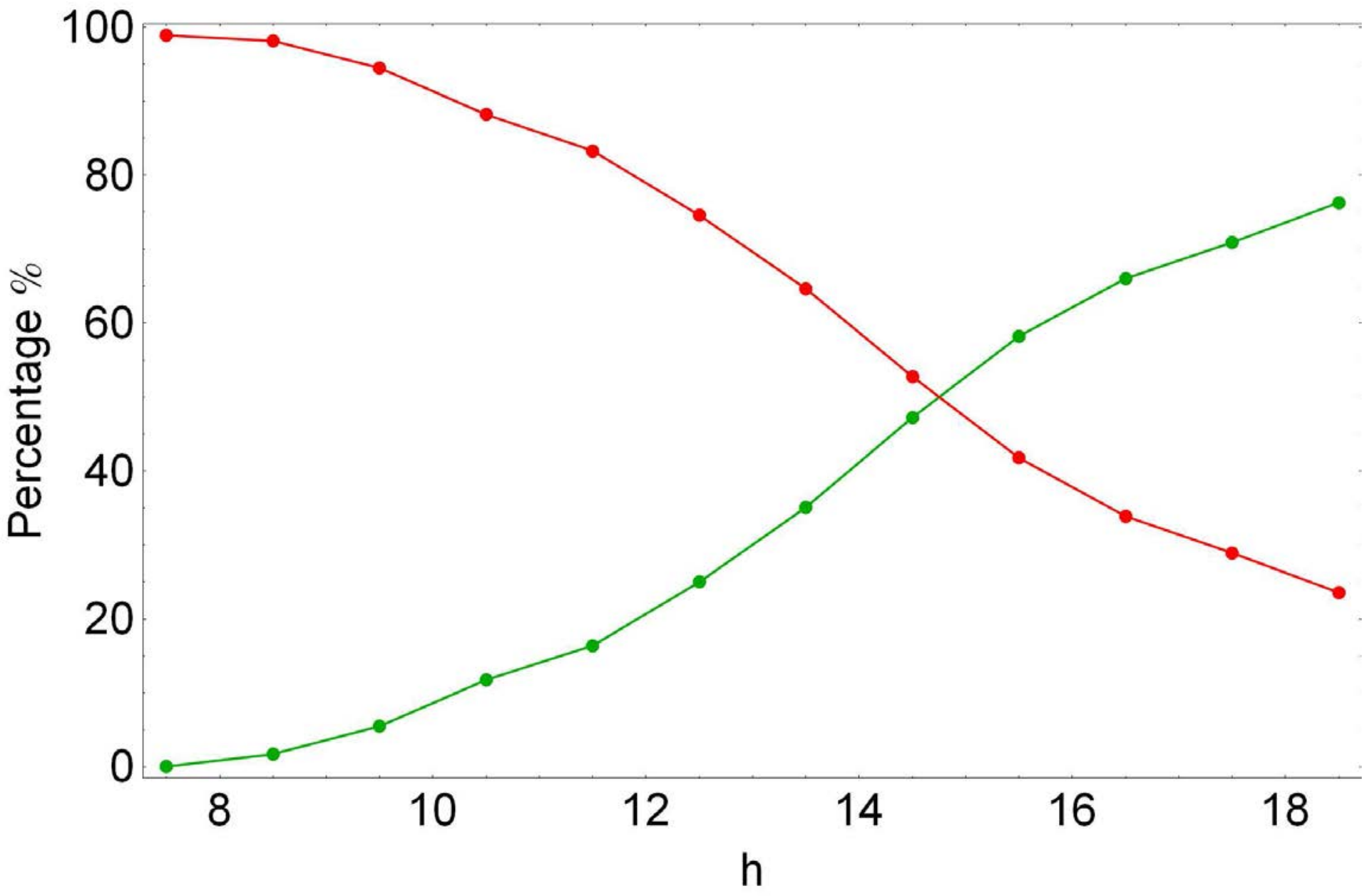}
\caption{Evolution of the percentages of regular escaping orbits (green) and chaotic escaping orbits (red) as a function of the energy $h$, when $z_0 = 0.1$.}
\label{percsz01}
\end{figure}

We start our exploration, considering orbits staring very close to the primary $(x,y)$ plane with initial condition $z_0 = 0.1$. Fig. \ref{rcz01} shows the structure of the $(x,\dot{x})$ plane for several values of the energy, where different colors indicate the four types of orbits. Specifically, black color corresponds to regular trapped orbits, blue color to chaotic trapped orbits, green color to regular escaping orbits, while the initial conditions of  chaotic escaping orbits are marked with red color. The outermost black solid line is the limiting curve defined by Eq. (\ref{zvc}). We observe, that for all studied energy levels the vast majority of the initial conditions correspond to either regular or chaotic escaping orbits, while trapped orbits are mainly confined to the center of the grid where small stability regions should exist. Here we should notice, that generally there is no indication of well-defined regions corresponding to a particular type of orbits. In fact, the initial conditions of the four types of orbits seem to be randomly scattered across the plane. It is also seen, that for low values of the energy (very close to the energy of escape) almost the entire grid is covered by chaotic escaping orbits. As the energy increases however, regular escaping orbits are gaining ground and finally they take over the grid. All these can be seen better in Fig. \ref{percsz01} where we present the evolution of the percentages of regular and chaotic escaping orbits as a function of the energy $h$. Indeed, the rates evolve similarly but following a complete different direction; chaotic escaping orbits decrease, while regular escaping orbits increase. It is worth noticing, that about $h = 15$ they share the whole area of the grid, while at the highest studied value of energy regular escaping orbits are four time more the chaotic escaping orbits\footnote{Generally, any dynamical method requires a sufficient time interval of numerical integration in order to distinguish safely between ordered and chaotic motion. Therefore, if the escape rate of orbits is very low or even worse if the orbits escape directly from the system then, any chaos indicator (the SALI in our case) will fail to work properly due to insufficient integration time. Nevertheless, we decided to apply the SALI method regardless of the escape rate of orbits.}.

\begin{figure*}
\centering
\resizebox{0.75\hsize}{!}{\includegraphics{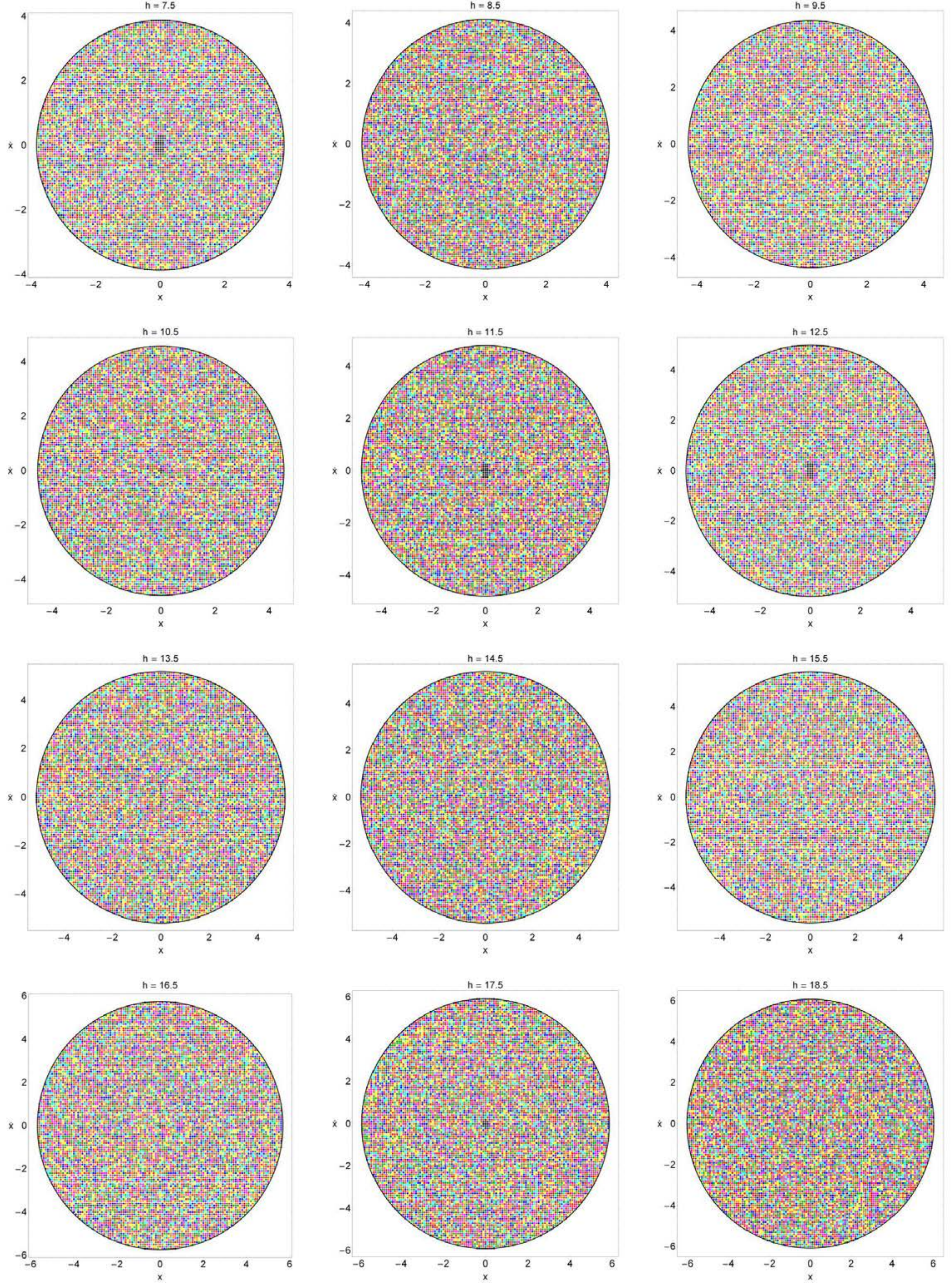}}
\caption{The structure of the $(x,\dot{x})$ plane for several values of the energy $h$, distinguishing between different escape channels, when $z_0 = 0.1$. The color code is the following: Trapped (black); channel 1 (red); channel 2 (green); channel 3 (brown); channel 4 (blue); channel 5 (orange); channel 6 (cyan); channel 7 (magenta); channel 8 (yellow).}
\label{chansz01}
\end{figure*}

\begin{figure}
\includegraphics[width=\hsize]{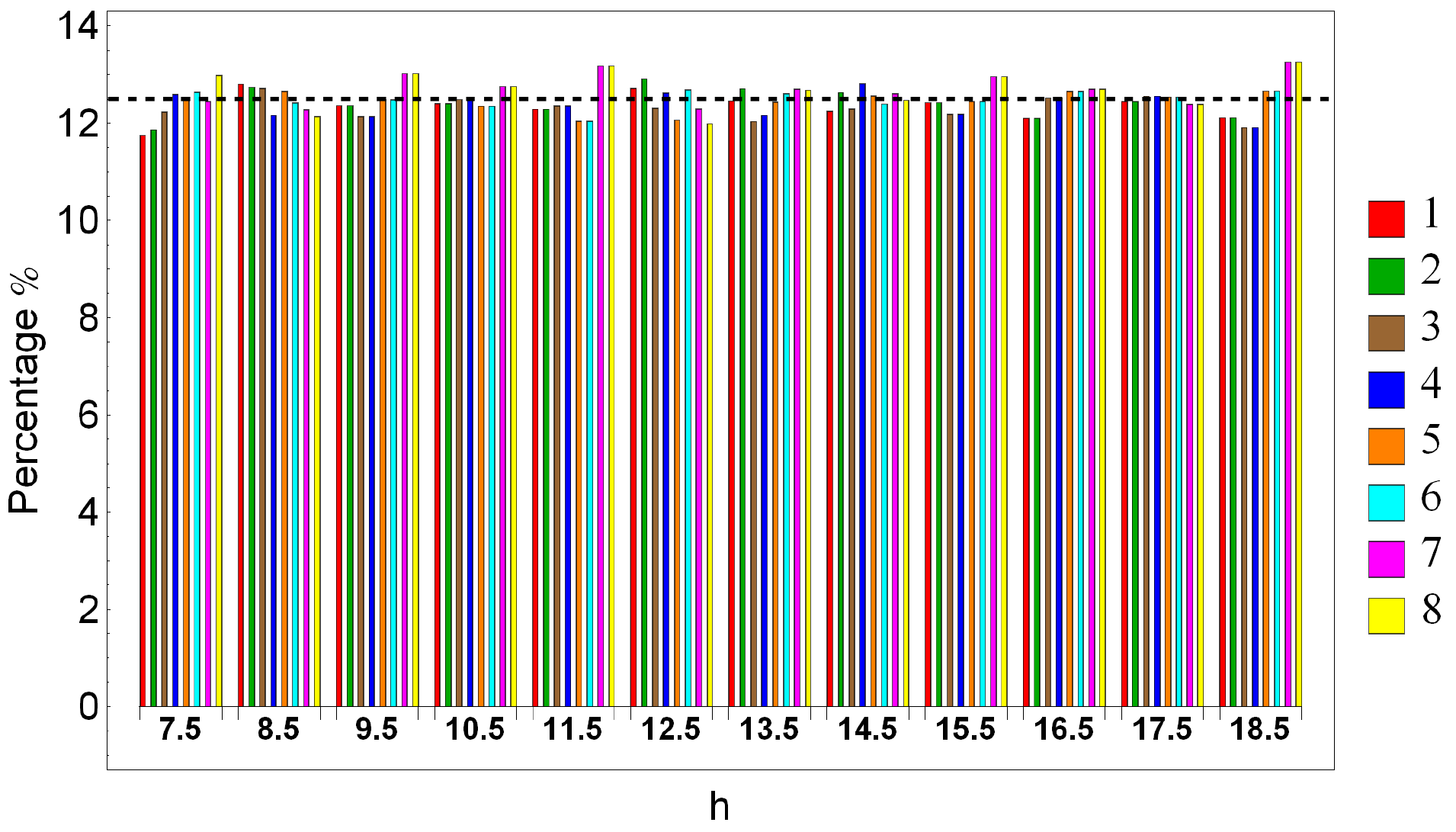}
\caption{A histogram of the percentages of escaping orbits per channel for all the tested values of the energy $h$, when $z_0 = 0.1$.}
\label{histz01}
\end{figure}

\begin{figure}
\includegraphics[width=\hsize]{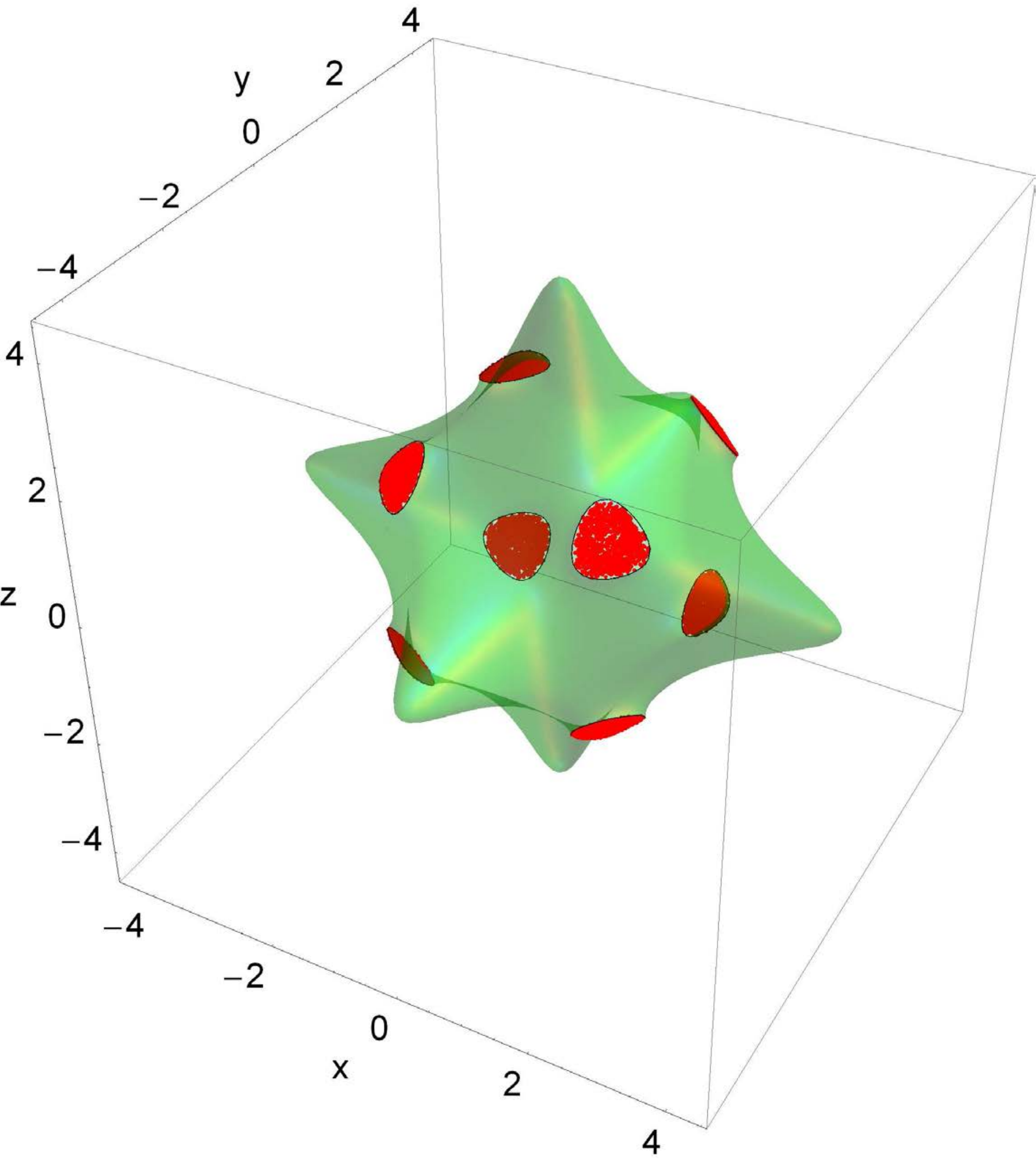}
\caption{The isopotential surface when $h = 8.5$, where the escape positions of the orbits are pinpointed with red dots.}
\label{exitsf}
\end{figure}

\begin{figure*}
\centering
\resizebox{0.8\hsize}{!}{\includegraphics{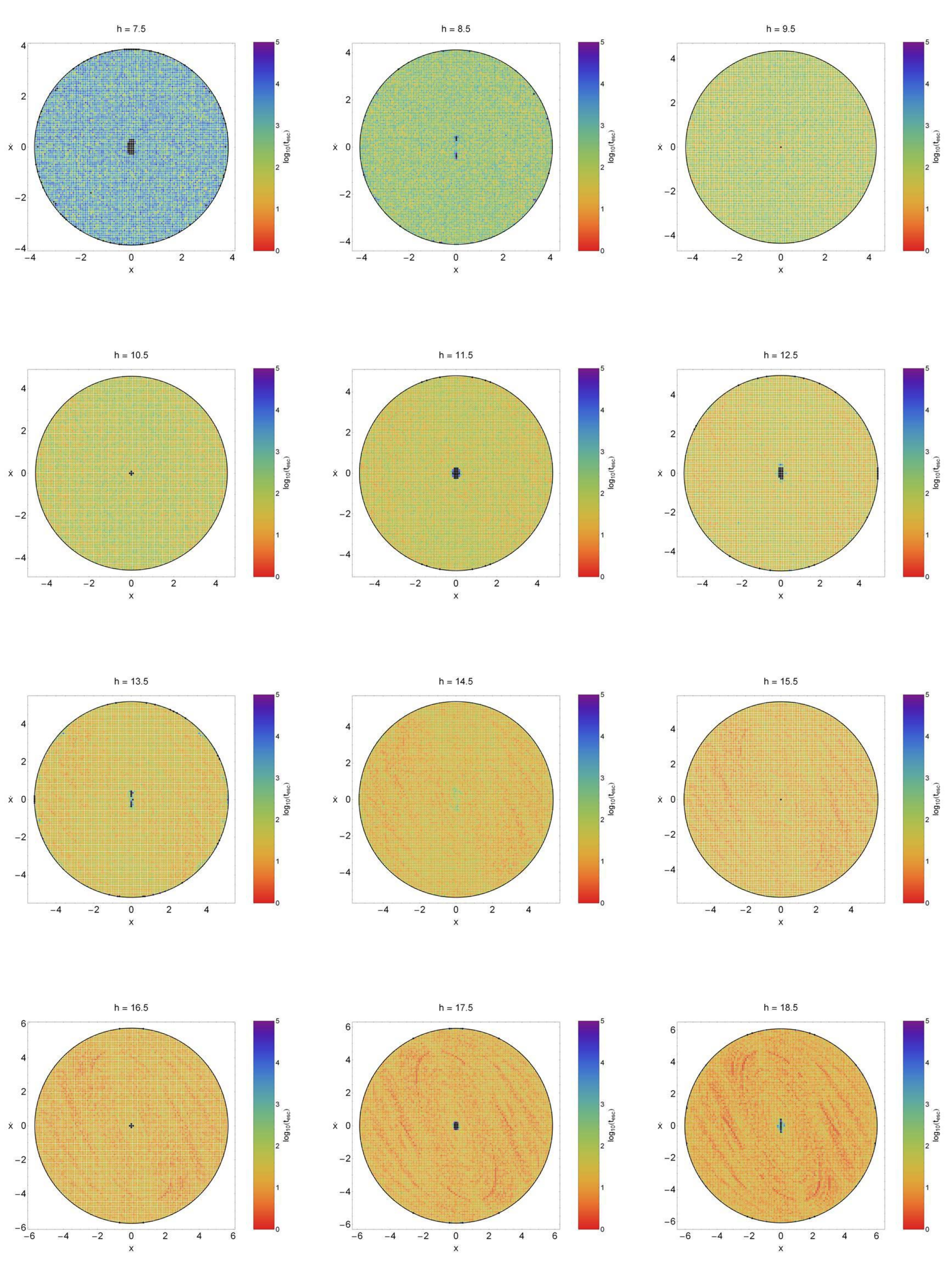}}
\caption{Distribution of the escape times $t_{\rm esc}$ of the orbits on the $(x,\dot{x})$ plane, when $z_0 = 0.1$. The darker the color, the larger the escape time. Trapped orbits are indicated by black color.}
\label{tescz01}
\end{figure*}

\begin{figure}
\includegraphics[width=\hsize]{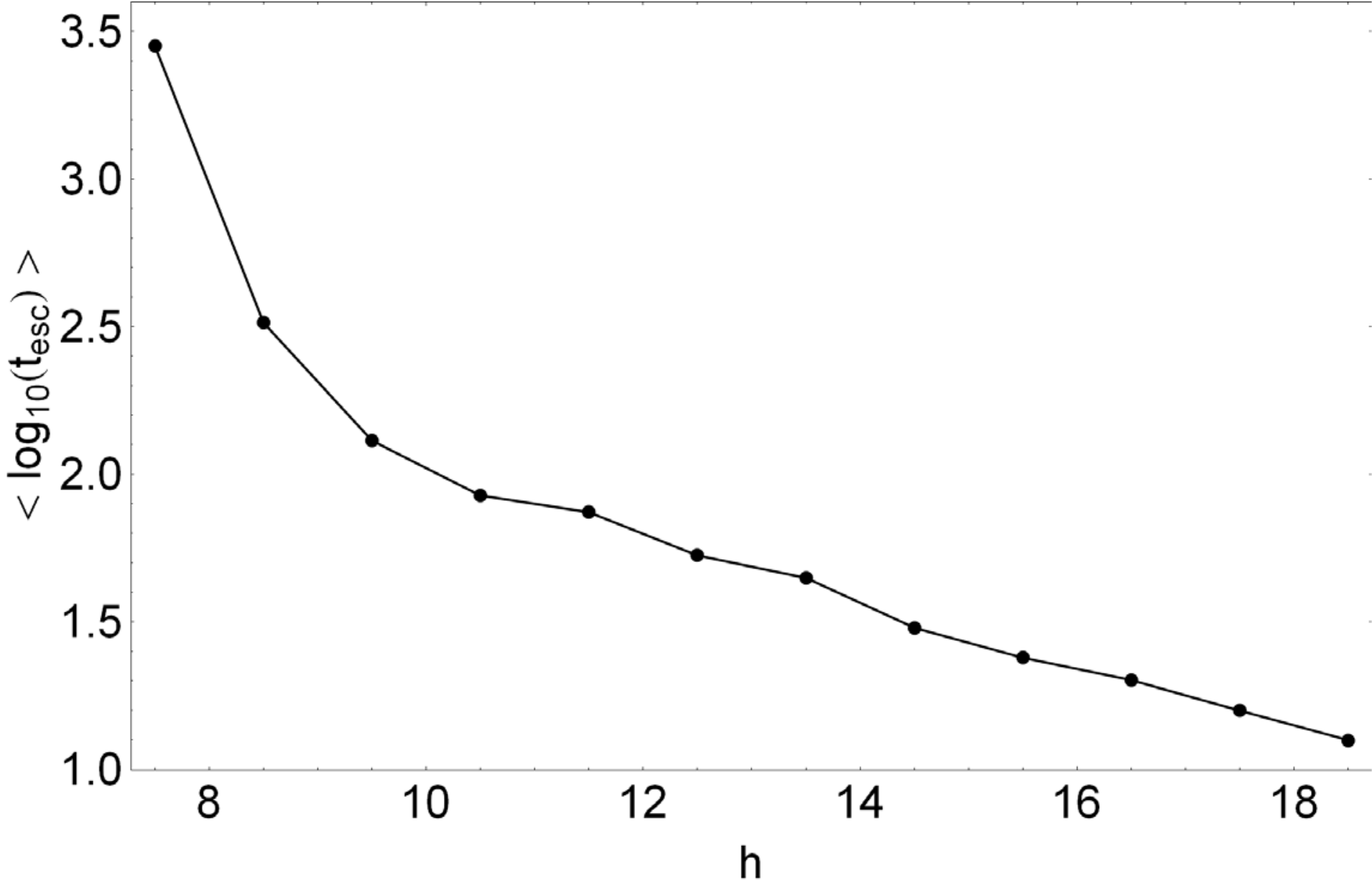}
\caption{Evolution of the average escape period of orbits as a function of the energy $h$, when $z_0 = 0.1$.}
\label{avgtz01}
\end{figure}

Fig. \ref{chansz01} depicts in another point of view the structure of the $(x,\dot{x})$ plane. In this case, each initial condition is colored according to the escape channel through which the particular orbit escapes. The black regions denote initial conditions where the test particles do not escape. Looking those grids we can draw two conclusions: (i) there is a highly sensitive dependence of the escape process on the initial conditions, that is, a slight change in the initial conditions makes the test particle escape through another channel, which is is a classical indication of chaos; (ii) All computed grids exhibit a very rich fractal structure regarding the escaping channels. This is true, because there is no indication whatsoever of basins of escape. By the term basin of escape, we refer to a set of initial conditions that corresponds to a certain escape channel. Inspecting carefully the grids shown in Fig. \ref{chansz01}, one may identify very small and local patterns leading to groups of neighbour initial conditions with the same color, however, these groups are extremely small and therefore cannot be recognized as basins of escape. The fractalization of the grid indicates that all channels of escape should be equiprobable. This is confirmed in Fig. \ref{histz01} where we present a histogram of the percentages of escaping orbits per channel for all the tested values of the energy. It is seen, that in general terms (in some cases, the rates of channels 7 and 8 are slightly elevated) all channels have the same percentage, roughly around 12.5\% (indicated by the horizontal black dashed line), which yields to equal probability of 1/8. In Fig. \ref{exitsf} we present the isopotential surface for $h = 8.5$, where we have marked with red dots the escape positions\footnote{The term ``escape positions" refers to the $(x,y,z)$ points of the 3D space at which the orbits intersect the cutoff surface with velocity pointing outwards and escape.} of the orbits. Taking into account that the corresponding grid contains a very rich sample of orbits (about 25000) and almost all of them escape choosing randomly and therefore equiprobably a channel, we see that all eight exits of the surface have been patched equally well thus verifying that in this case the escape channels are indeed equiprobable.

The following Fig. \ref{tescz01} shows how the escape times $t_{\rm esc}$ of orbits are distributed on the grid. Light reddish colors correspond to fast escaping orbits, dark blue/purpe colors indicate large escape periods, while black color denote trapped orbits. We observe, that when $h = 7.5$, that is a value of energy very close to the escape energy, the escape periods of the majority of orbits are huge corresponding to tens of thousands of time units. This however, is anticipated because in this case the width of the escape channels is very small and therefore, the orbits should spend much time inside the isopotential surface until they find one of the openings. As the value of the energy increases, the escape channels become more and more wide leading to faster escaping orbits, which means that the escape period decreases rapidly. To prove this point, we calculated for each value of the energy the respective average value of the escape period $< t_{\rm esc} >$. Our results are given in Fig. \ref{avgtz01}, where we see that for $h < 10$ the escape period of orbits decreases rapidly, while for larger energies it follows an almost linear decrease. Here we should note, that for large values of the energy $(h > 16.5)$ weak color patterns of elongated bands spiralling around the center of the grids appear. These patterns correspond to extremely fast escaping orbits with escape periods less than 10 time units.

\subsection{Case II: Orbits starting with a mediocre value of $z_0$}
\label{case2}

\begin{figure*}
\centering
\resizebox{0.8\hsize}{!}{\includegraphics{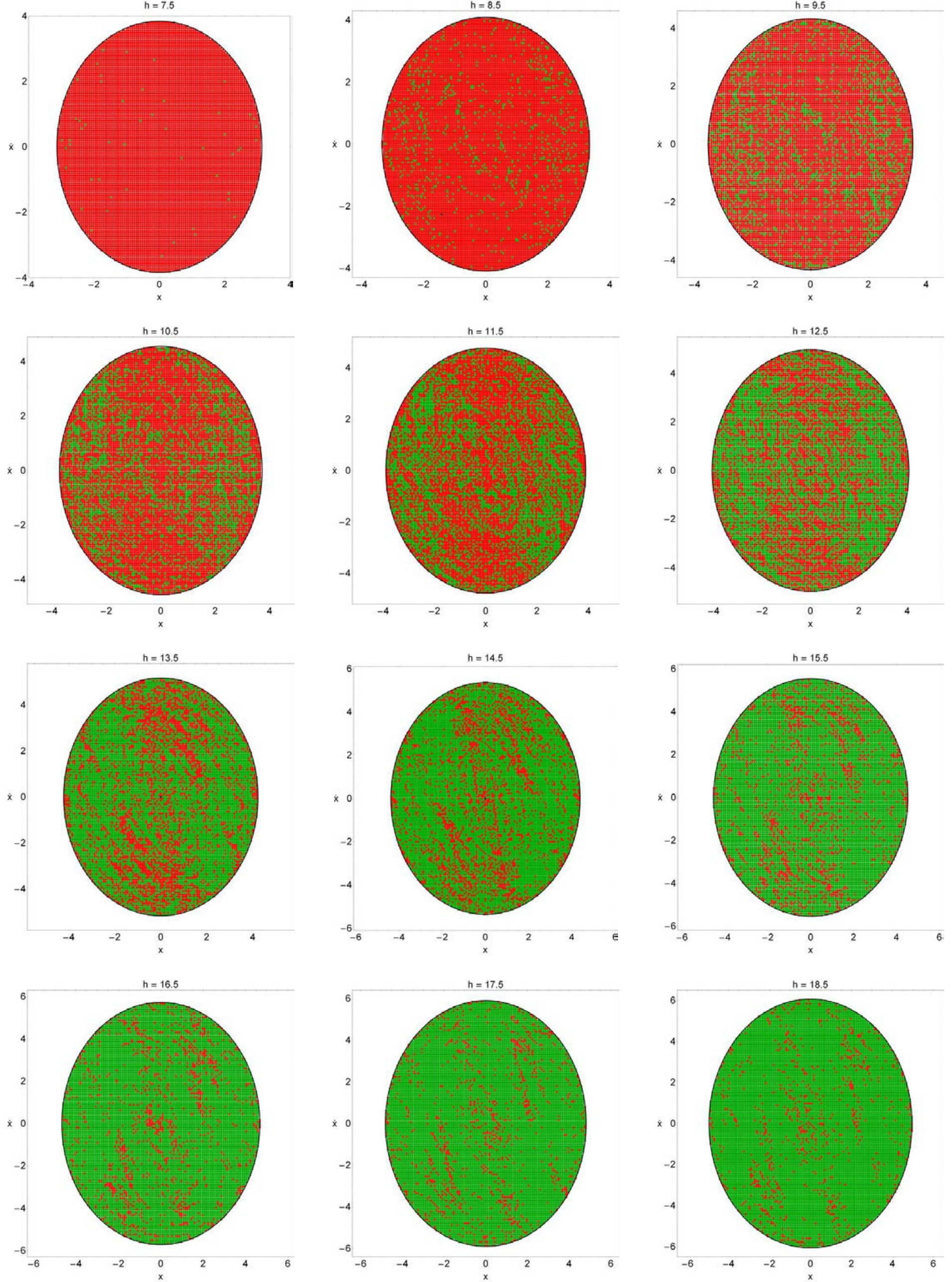}}
\caption{The structure of the $(x,\dot{x})$ plane for several values of the energy $h$, distinguishing between regular trapped orbits (black), chaotic trapped orbits (blue), regular escaping (green) and chaotic escaping (red), when $z_0 = 0.5$.}
\label{rcz05}
\end{figure*}

\begin{figure}
\includegraphics[width=\hsize]{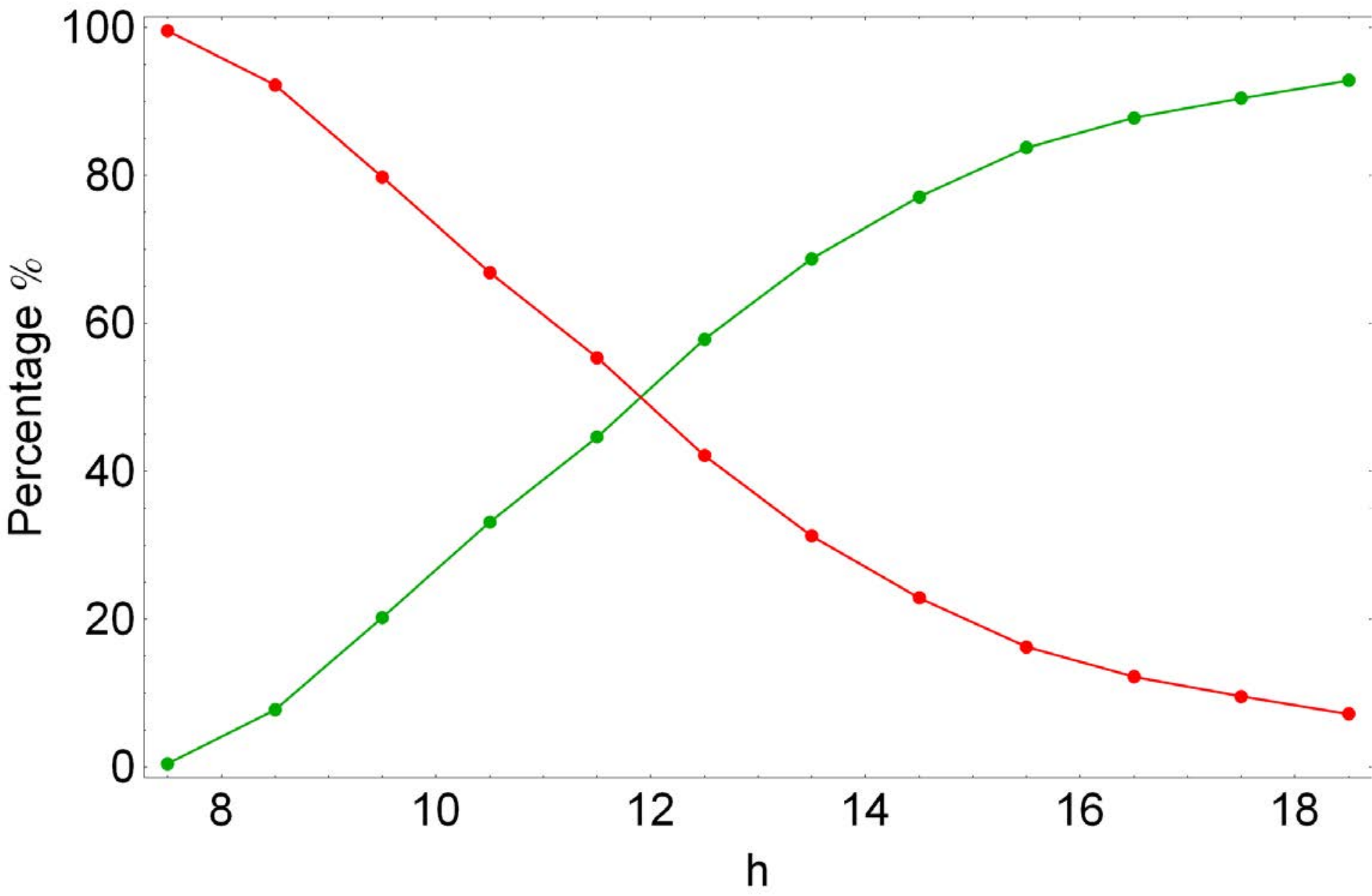}
\caption{Evolution of the percentages of regular escaping orbits (green) and chaotic escaping orbits (red) as a function of the energy $h$, when $z_0 = 0.5$.}
\label{percsz05}
\end{figure}

Our investigation continues considering in this case orbits starting at a moderate distance from the primary $(x,y)$ plane having initial condition $z_0 = 0.5$. We shall follow the same numerical approach as in Case I. The structure of the $(x,\dot{x})$ plane using different colors in order to distinguish between the four types of orbits (regular trapped; chaotic trapped; regular escaping and chaotic escaping) is presented in Fig. \ref{rcz05}. We observe similar results to that discussed previously in Fig. \ref{rcz01}. It is worth noticing, that the increase of the initial value of the $z$ coordinate of orbits, has affected the geometry of the grid since it has become more oval. Our numerical calculations reveal, that in most examined cases all the integrated orbits of the grids escape sooner or later, while only the $(0,0)$ periodic orbits do not escape and only when they are stable\footnote{The stability of a periodic orbit is determined by computing the stability index [\citealp{Z13}]. A periodic orbits is stable if only the stability index (S.I.) is between -2 and +2.}. We also see, that for low values of the energy chaotic escaping orbits swarm the grid, while as the energy increases regular escaping orbits become the most populated type of orbits. Fig. \ref{percsz05} shows how the energy influence the percentages of regular and chaotic escaping orbits. The pattern of the evolution is quite similar to that of Fig. \ref{percsz01} however, there are two noticeable differences: (i) the point where regular and chaotic escaping orbits share the grid comes earlier at about $h = 12$ and (ii) at the highest studied energy level regular escaping orbits are nine time more the chaotic escaping orbits.

\begin{figure*}
\centering
\resizebox{0.75\hsize}{!}{\includegraphics{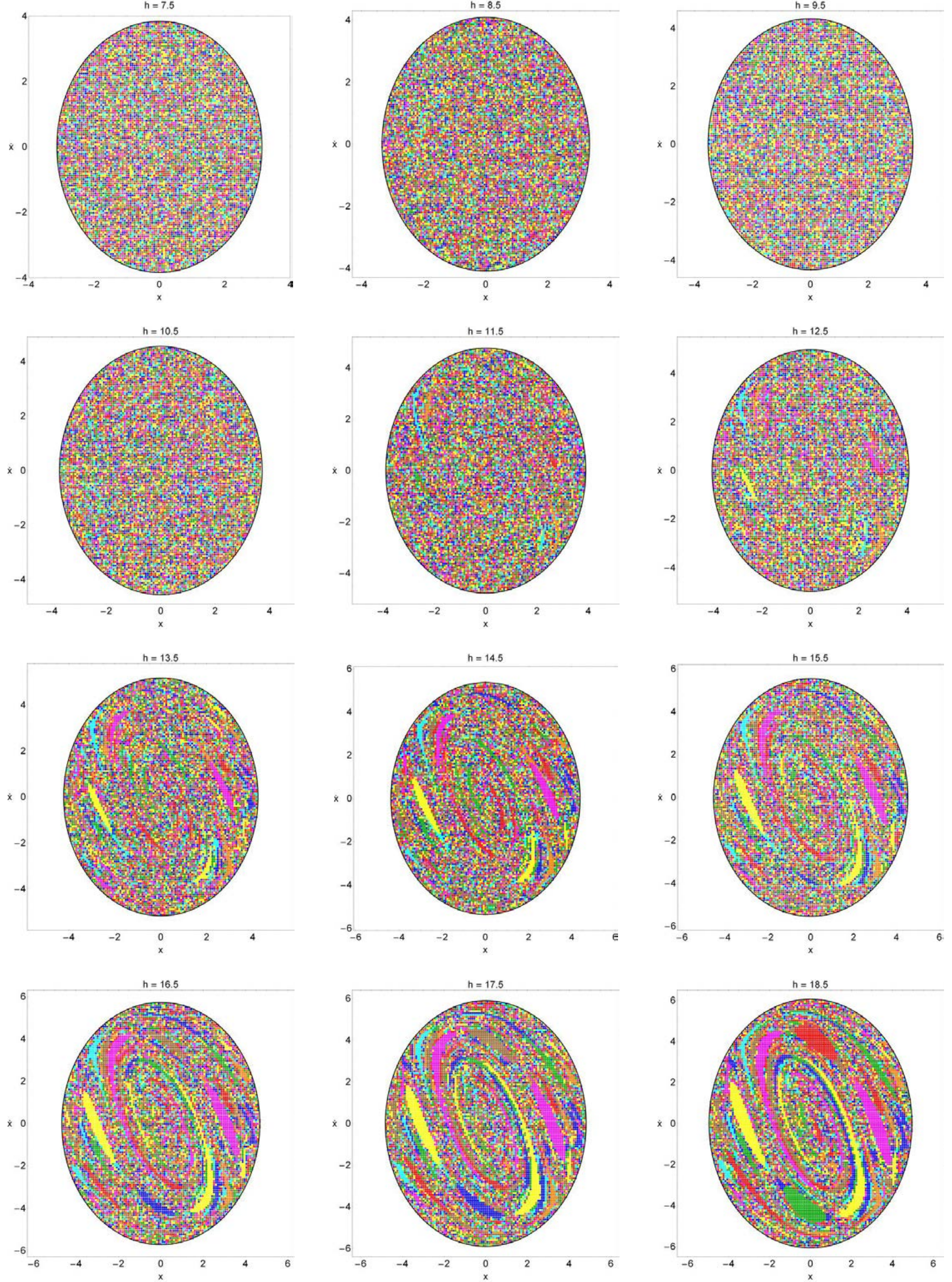}}
\caption{The structure of the $(x,\dot{x})$ plane for several values of the energy $h$, distinguishing between different escape channels, when $z_0 = 0.5$. The color code is the following: Trapped (black); channel 1 (red); channel 2 (green); channel 3 (brown); channel 4 (blue); channel 5 (orange); channel 6 (cyan); channel 7 (magenta); channel 8 (yellow).}
\label{chansz05}
\end{figure*}

\begin{figure}
\includegraphics[width=\hsize]{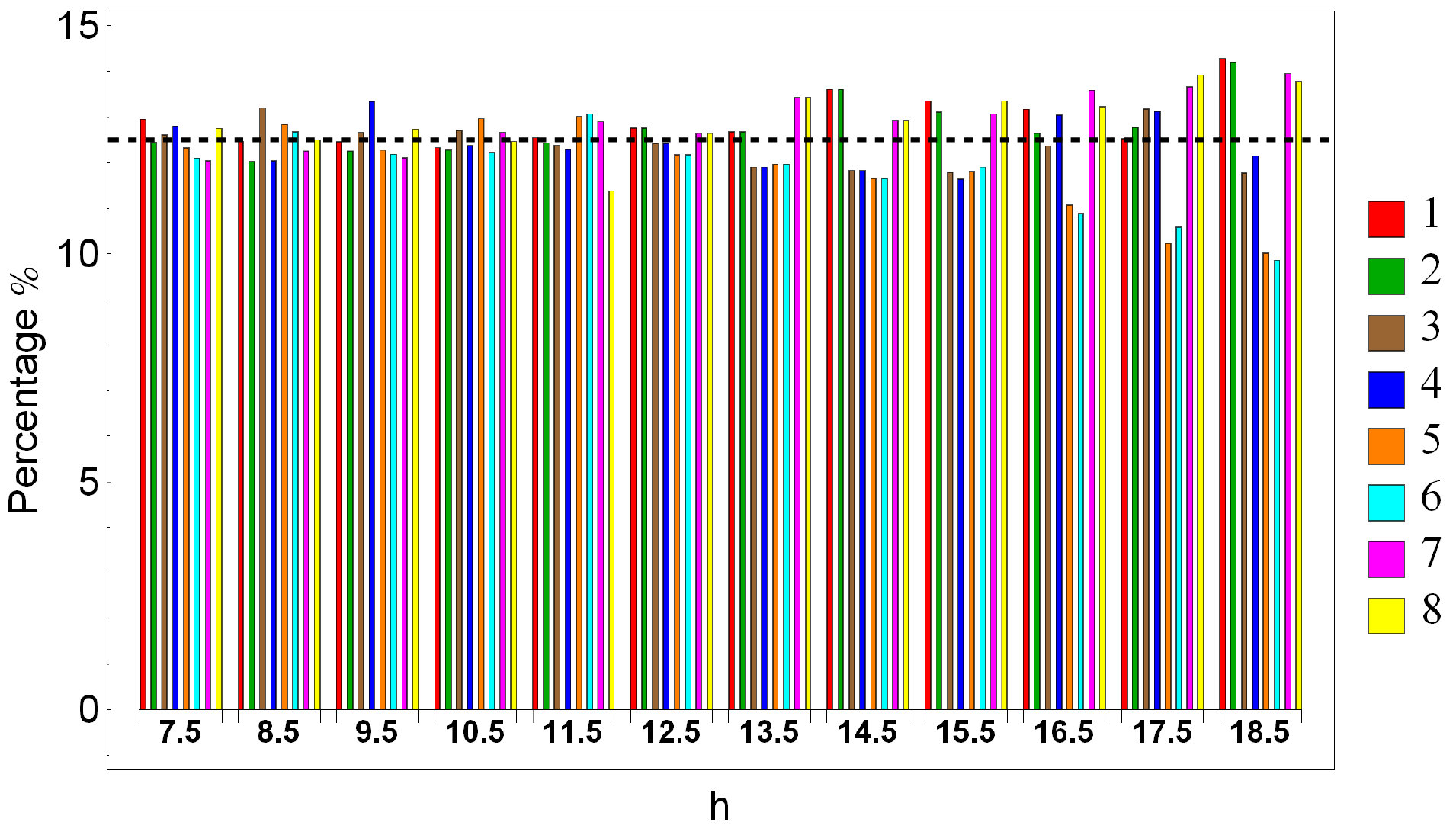}
\caption{A histogram of the percentages of escaping orbits per channel for all the tested values of the energy $h$, when $z_0 = 0.5$.}
\label{histz05}
\end{figure}

\begin{figure*}
\centering
\resizebox{0.8\hsize}{!}{\includegraphics{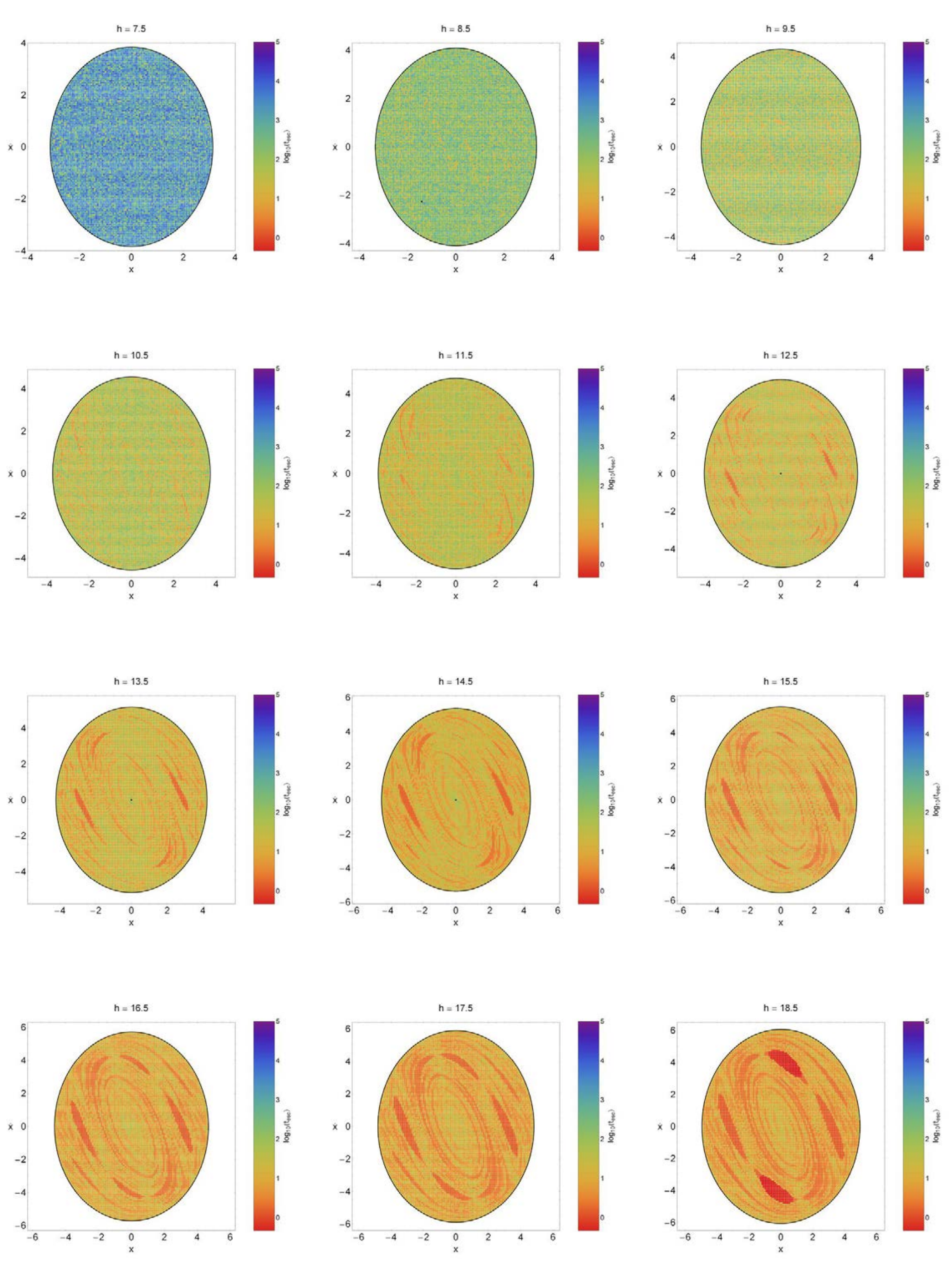}}
\caption{Distribution of the escape times $t_{\rm esc}$ of the orbits on the $(x,\dot{x})$ plane, when $z_0 = 0.5$. The darker the color, the larger the escape time. Trapped orbits are indicated by black color.}
\label{tescz05}
\end{figure*}

\begin{figure}
\includegraphics[width=\hsize]{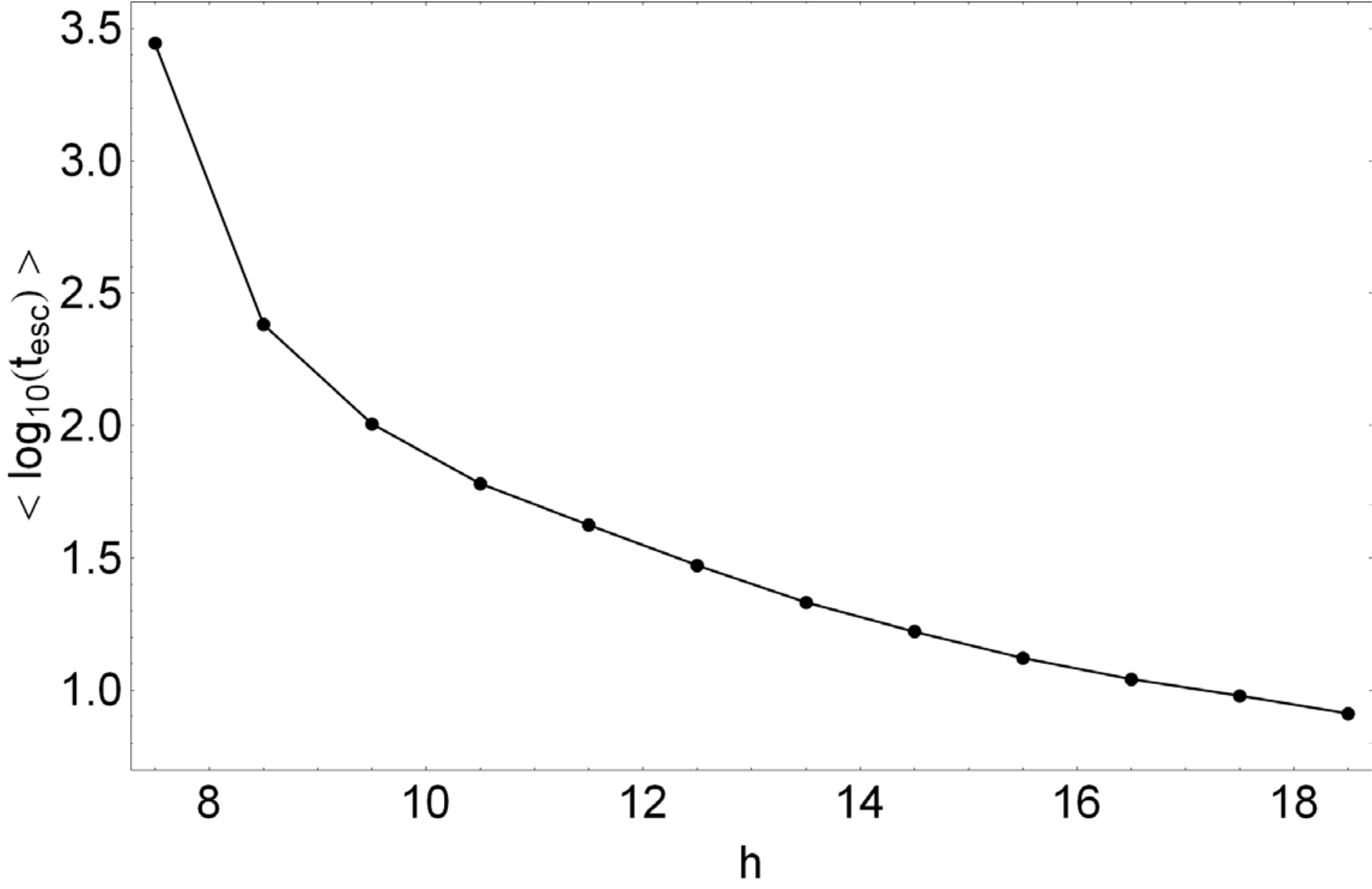}
\caption{Evolution of the average escape period of orbits as a function of the energy $h$, when $z_0 = 0.5$.}
\label{avgtz05}
\end{figure}

The $(x,\dot{x})$ plane is reconstructed in Fig. \ref{chansz05}, where each initial condition is colored according to the escape channel through which the particular orbit escapes. It is seen, that for low values of the energy the grids exhibit a large degree of fractalization regarding the escape channels. However, when $h > 11$ small groups of neighbor initial conditions leading to the same exit start to emerge. As the energy increases, these structures expand in size thus forming well-defined basins of escape, while the degree of fractalization of the grid is consequently reduced. The basins of escape are either broad regions, or elongated bands spiralling around the center of the grid. The various basins of escape are very intricately interwoven. The formation of several basins of escape implies that the escape channels should \emph{not} be equiprobable any longer. This assumption is justified in Fig. \ref{histz05}, where a histogram of the percentages of escaping orbits per channel for all the tested values of the energy is given. One can observe, that when $h > 12.5$ the test particles seem not to choose randomly an exit but they prefer specific channels through which they escape. In general terms, channels 1, 2, 7 and 8 appear to be the most favorite ones and are more likely to be chosen, thus having a probability higher than the average. Exits 3 and 6 on the other hand, are the least favorite having the smallest probability to be chosen.

The distribution of the escape times of orbits on the grid when $z_0 = 0.5$ is shown in Fig. \ref{tescz05}. We observe several similarities with Fig. \ref{tescz01} which discussed in the previous case where $z_0 = 0.1$ being the most noticeable the large escape periods of orbits for values of the energy very close to the escape energy and the rapid decrease of the escape times with increasing energy. This behavior can be comprehended better by looking at Fig. \ref{avgtz05} where the evolution of the average value of the escape period $< t_{\rm esc} >$ of orbits as a function of the energy is given. We see, that the average escape time displays a rapid reduction when $h < 9 $, while for larger energy values it decreases almost linearly. At this point, we should notice that the basins of escape for $h > 11$, can be distinguished in the grids of Fig. \ref{tescz05}, being the regions with intense red color indicating extremely fast escaping orbits. Our numerical calculations indicate, that orbits with initial conditions inside the basins have significantly small escape periods of less than 5 time units.

\subsection{Case III: Orbits starting with a high value of $z_0$}
\label{case3}

\begin{figure*}
\centering
\resizebox{0.8\hsize}{!}{\includegraphics{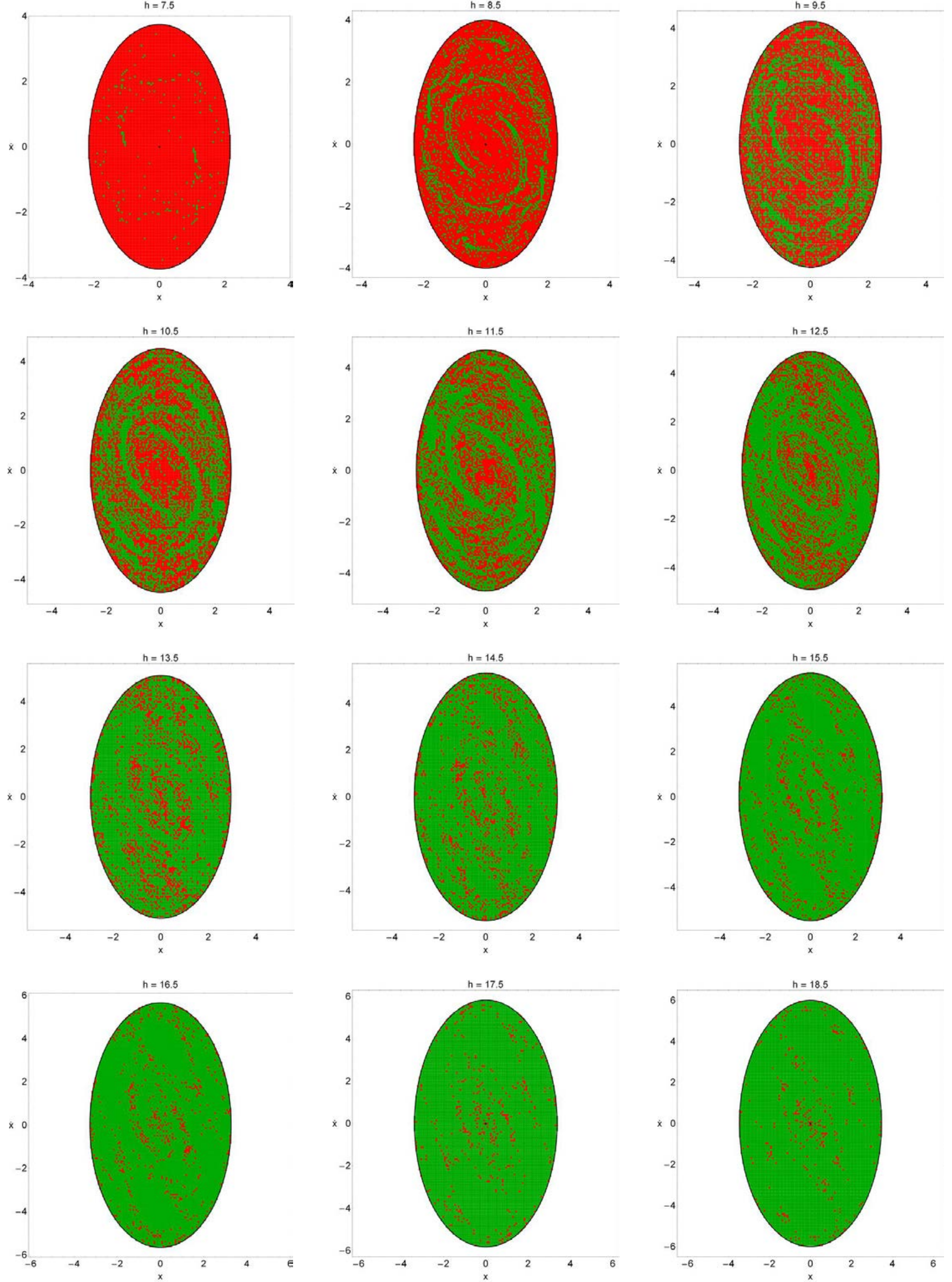}}
\caption{The structure of the $(x,\dot{x})$ plane for several values of the energy $h$, distinguishing between regular trapped orbits (black), chaotic trapped orbits (blue), regular escaping (green) and chaotic escaping (red), when $z_0 = 1$.}
\label{rcz10}
\end{figure*}

\begin{figure}
\includegraphics[width=\hsize]{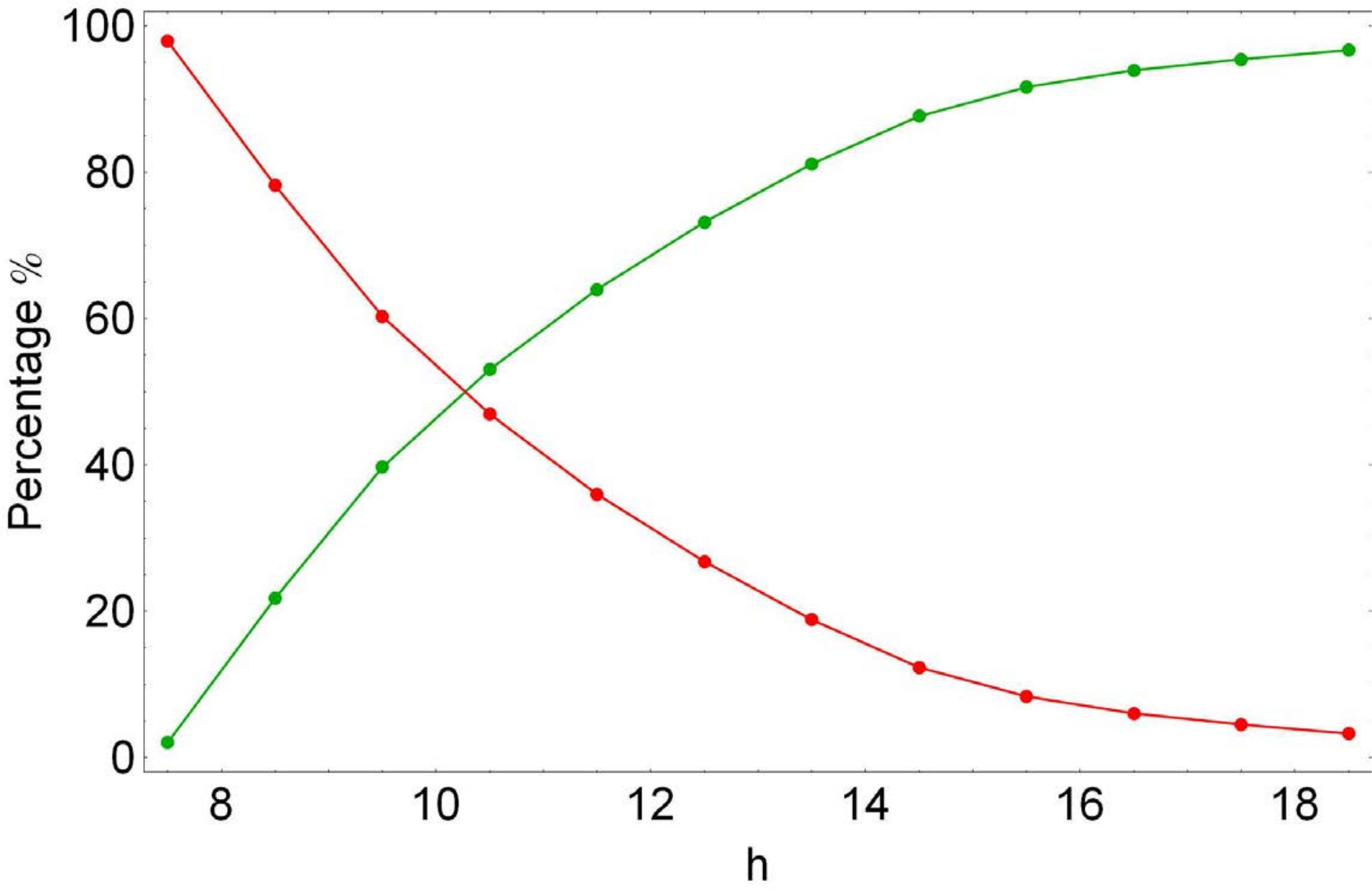}
\caption{Evolution of the percentages of regular escaping orbits (green) and chaotic escaping orbits (red) as a function of the energy $h$, when $z_0 = 1$.}
\label{percsz10}
\end{figure}

The last case under investigation involves orbits launched at a relatively high distance from the $(x,y)$ plane having initial condition $z_0 = 1$. Again, the numerical approach remains the same as in the previously studied cases. Fig. \ref{rcz10} shows the structure of the $(x,\dot{x})$ plane distinguishing between ordered/chaotic and trapped/escaping orbits using different colors. We see that things are quite similar to those of Figs. \ref{rcz01} and \ref{rcz05} however, the shape of the grid, or in other words the liming curve, has become much more oval and elongated along the $\dot{x}$ axis, due to the further increase of the initial value of the $z$ coordinate. As in the previous cases, only the central periodic orbit at $(0,0)$ remains trapped when it is stable, while all the other orbits escape from the system. Once more it is evident, that at low values of the energy the majority of escaping orbits are chaotic, while at high values of the energy regular escaping orbits dominate the grid. The evolution of the percentages of regular and chaotic escaping orbits as a function of the energy is given in Fig. \ref{percsz10}. It can be seen, that at about $h = 10$ regular and chaotic escaping orbit share the entire grid, while at high energy levels $h > 16.5$, chaotic escaping orbits are depleted and more than 95\% of the plane is covered by fast regular escaping orbits.

\begin{figure*}
\centering
\resizebox{0.75\hsize}{!}{\includegraphics{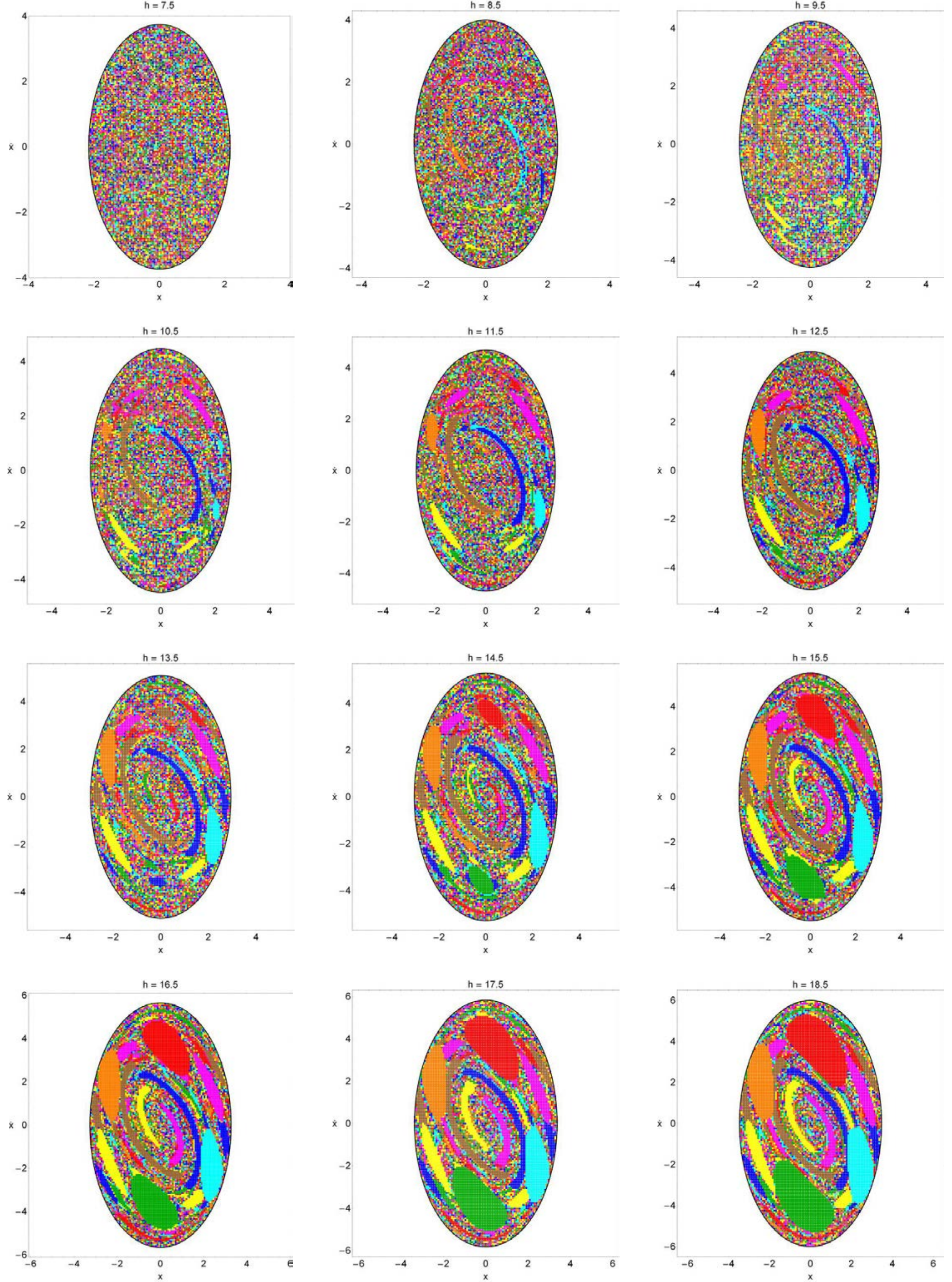}}
\caption{The structure of the $(x,\dot{x})$ plane for several values of the energy $h$, distinguishing between different escape channels, when $z_0 = 1$. The color code is the following: Trapped (black); channel 1 (red); channel 2 (green); channel 3 (brown); channel 4 (blue); channel 5 (orange); channel 6 (cyan); channel 7 (magenta); channel 8 (yellow).}
\label{chansz10}
\end{figure*}

\begin{figure}
\includegraphics[width=\hsize]{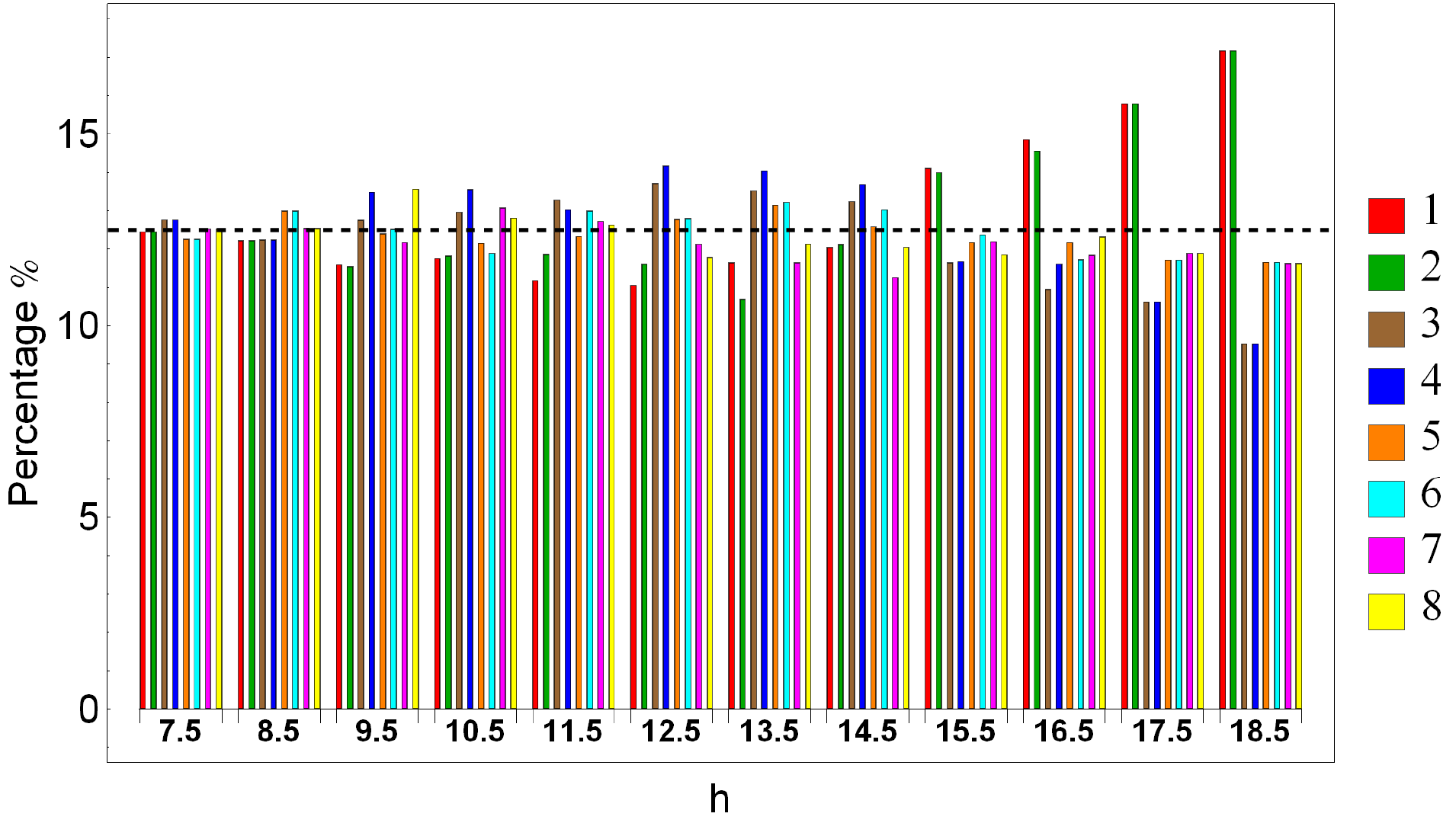}
\caption{A histogram of the percentages of escaping orbits per channel for all the tested values of the energy $h$, when $z_0 = 1$.}
\label{histz10}
\end{figure}

\begin{figure*}
\centering
\resizebox{0.8\hsize}{!}{\includegraphics{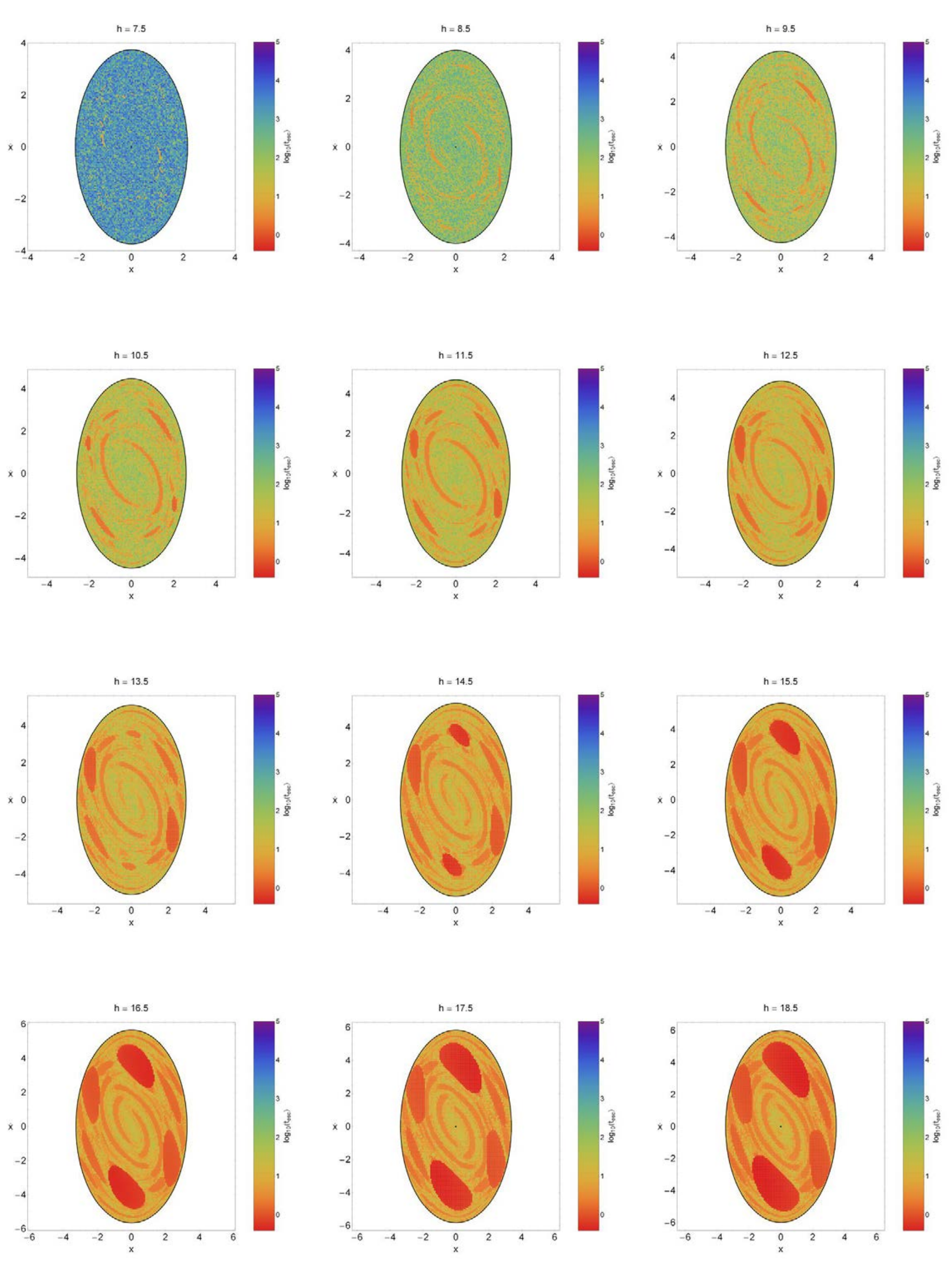}}
\caption{Distribution of the escape times $t_{\rm esc}$ of the orbits on the $(x,\dot{x})$ plane, when $z_0 = 1$. The darker the color, the larger the escape time. Trapped orbits are indicated by black color.}
\label{tescz10}
\end{figure*}

\begin{figure}
\includegraphics[width=\hsize]{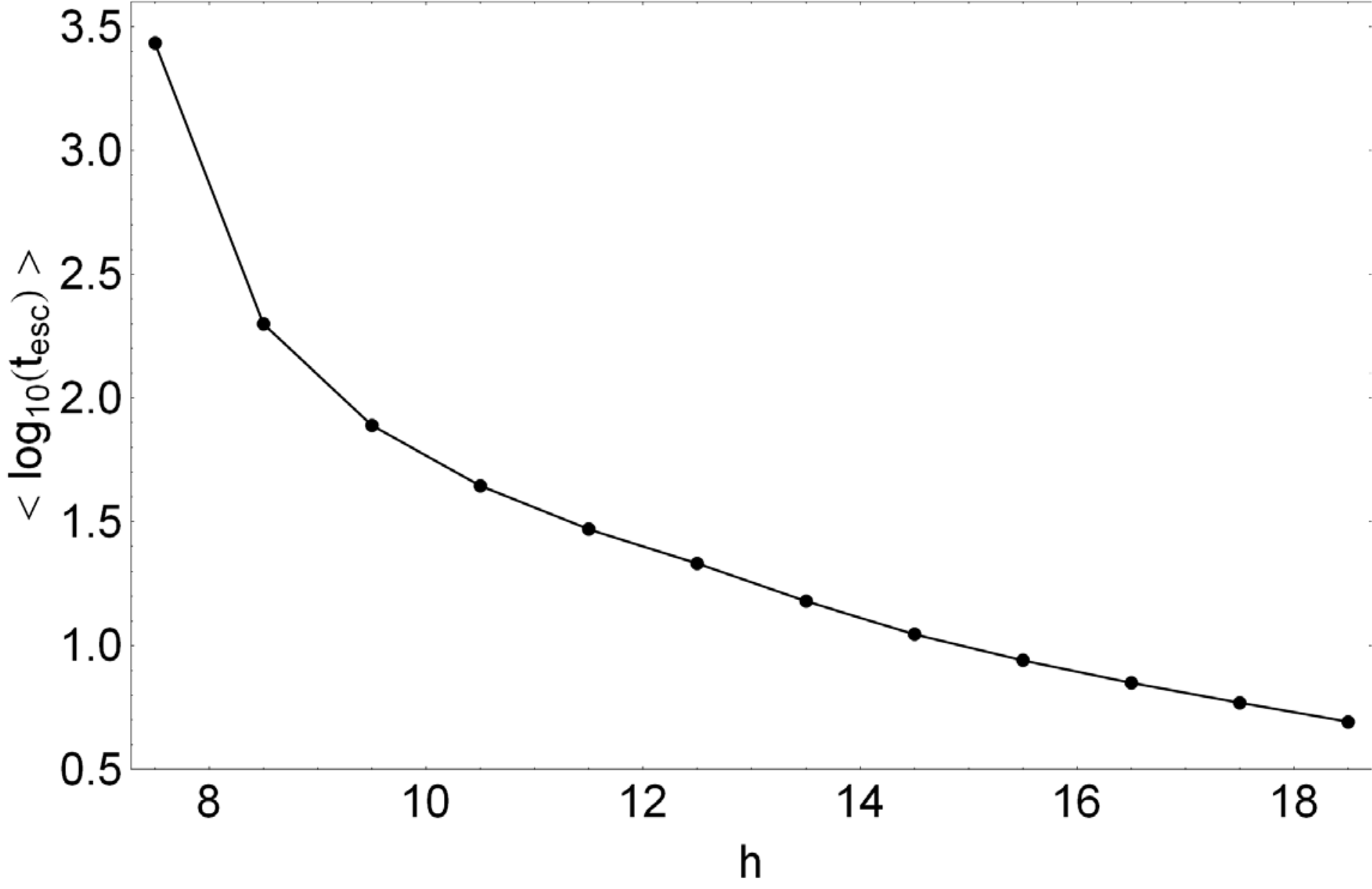}
\caption{Evolution of the average escape period of orbits as a function of the energy $h$, when $z_0 = 1$.}
\label{avgtz10}
\end{figure}

In Fig. \ref{chansz10} we present another point of view of the grids shown in Fig. \ref{rcz10}. Now, the color code of the initial conditions of orbits depends on the escape channels. One may observe, that apart from the $h = 7.5$ gird in which the structure of the $(x,\dot{x})$ plane displays high fractalization, in all other studied cases several basins of escape are present. At low values of the energy, we observe that the basins of escape appear only as weak elongated bands spiralling around the center, while as we proceed to grids corresponding to higher energy levels broad escape regions emerge which eventually take over the majority of the grid at high energies $(h > 15)$. Therefore, we may conclude that the increase of the energy favors the growth and the development of basins of escape. Fig. \ref{histz10} shows the percentages of escaping orbits per channel for all the tested values of the energy. Evidently, the escape channels can be considered equiprobable only at values of energy very close to the escape energy $(h < 9)$. In particular, for $9.5 \leq h \leq 14.5$ channels 3 and 4 seem to be the most favorable, while this behavior is reversed for larger values of the energy. Indeed, we observe that when $h > 15$ channels 1 and 2 are clearly the most selected ones, while channels 3 and 4 are the least chosen.

\begin{figure}
\includegraphics[width=\hsize]{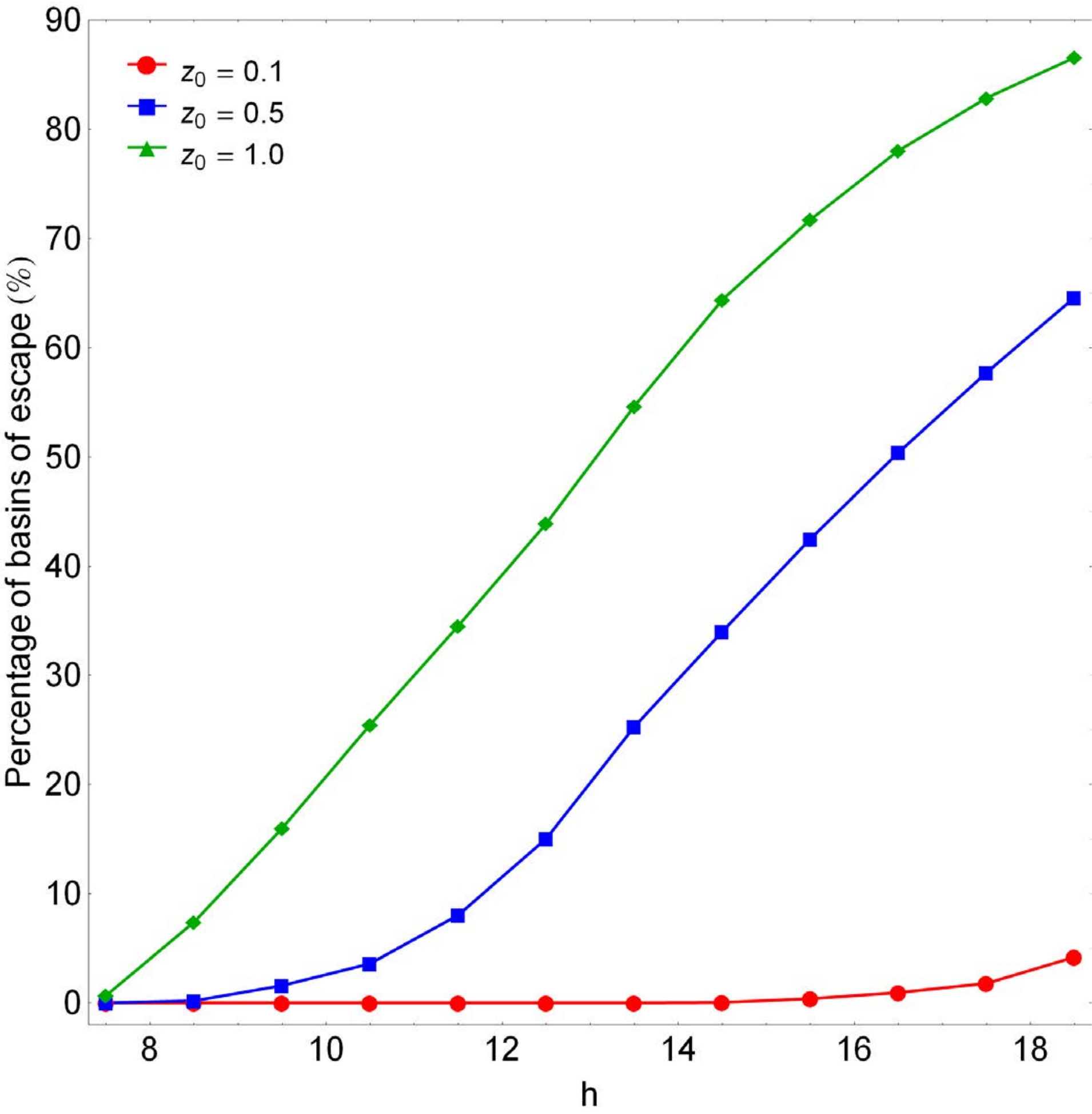}
\caption{Evolution of the percentage of the total area on the $(x,\dot{x})$ plane corresponding to basins of escape, as a function of the energy $h$ for the three initial values of the $z$ coordinate.}
\label{basins}
\end{figure}

Finally, Fig. \ref{tescz10} shows the distribution of the escape times of orbits on the grids when $z_0 = 1$. The results are very similar to those presented in Figs. \ref{tescz01} and \ref{tescz05}. We see, that the general rule according to which the escape times decrease with increasing energy is still valid. The evolution of the average value of the escape period $< t_{\rm esc} >$ of orbits as a function of the energy given in Fig. \ref{avgtz10} is also in the same vein. As we pointed out in the previous case, the several basins of escape are visible in the grids of Fig. \ref{tescz10}, being the regions with intense red color indicating orbits with extremely low escape periods. According to our numerical computations, orbits with initial conditions $(x_0,\dot{x_0})$ inside the basins escape almost immediately from the system.

\subsection{General overview}
\label{geno}

\begin{figure*}
\centering
\resizebox{0.8\hsize}{!}{\includegraphics{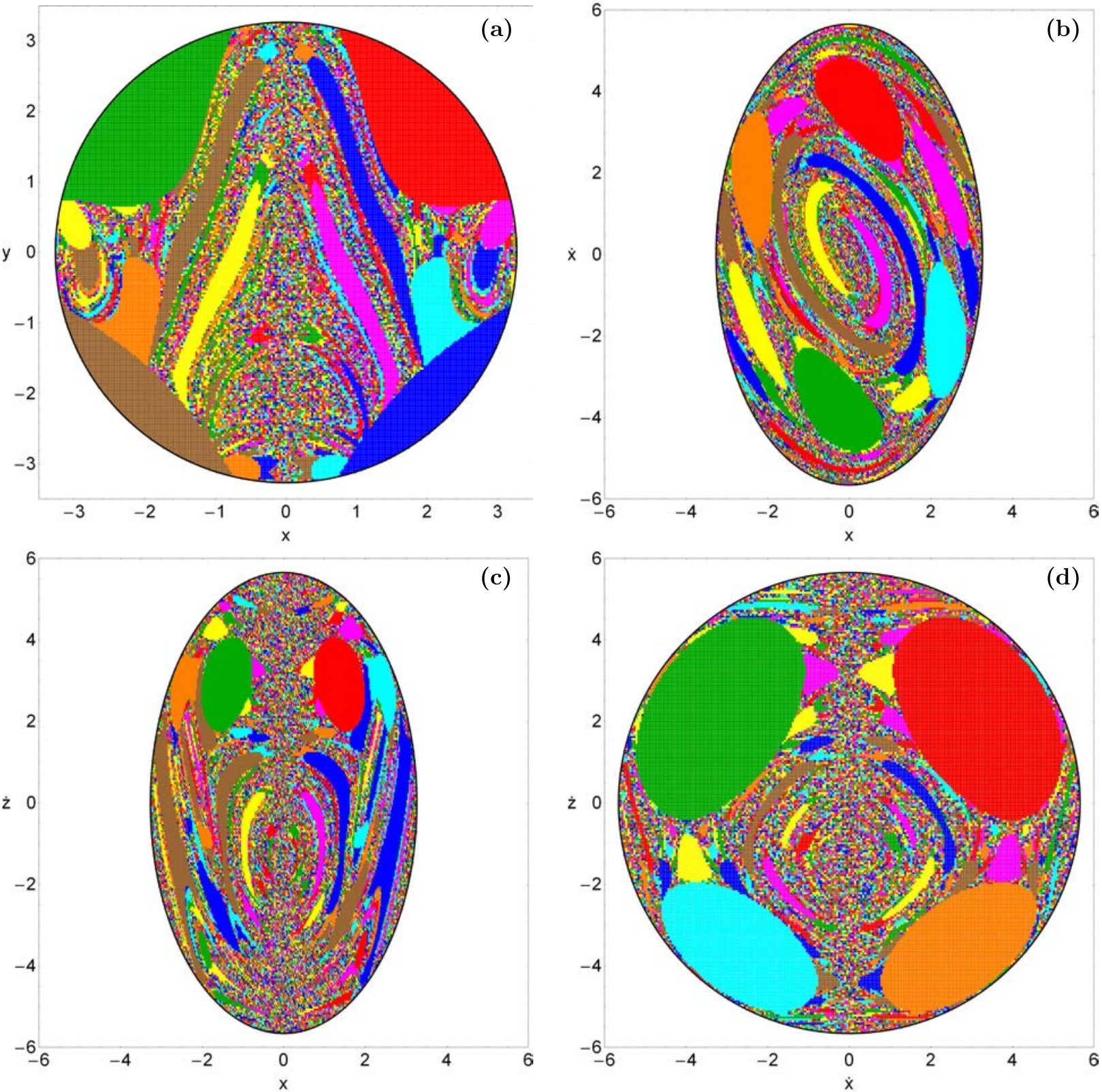}}
\caption{Grids of initial conditions in different two-dimensional sub-planes of the entire six-dimensional phase space for the $(h = 16.5, z_0 = 1)$ model. (a-upper left): $(x,y$ plane, (b-upper right): $(x,\dot{x})$ plane, (c-lower left): $(x,\dot{z})$ plane and (d-lower right): $(\dot{x},\dot{z})$ plane.}
\label{gridsm}
\end{figure*}

\begin{figure}
\includegraphics[width=\hsize]{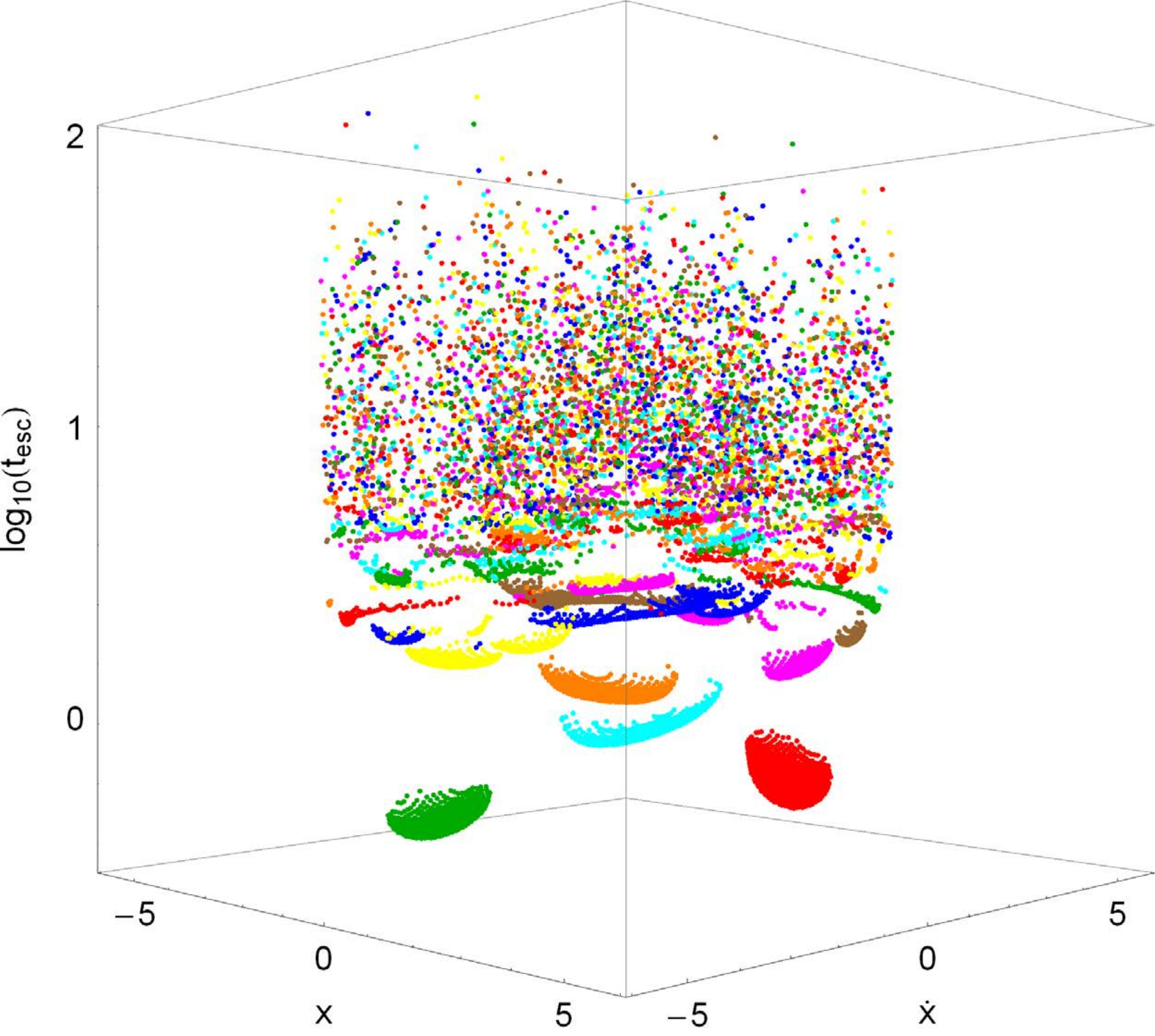}
\caption{A combination of the escape channels and the escape periods of orbits of the $(h = 16.5, z_0 = 1)$ model. The color code regarding the escape channels is the same as in Fig. \ref{chansz10}.}
\label{bas3d}
\end{figure}

\begin{figure*}
\centering
\resizebox{0.8\hsize}{!}{\includegraphics{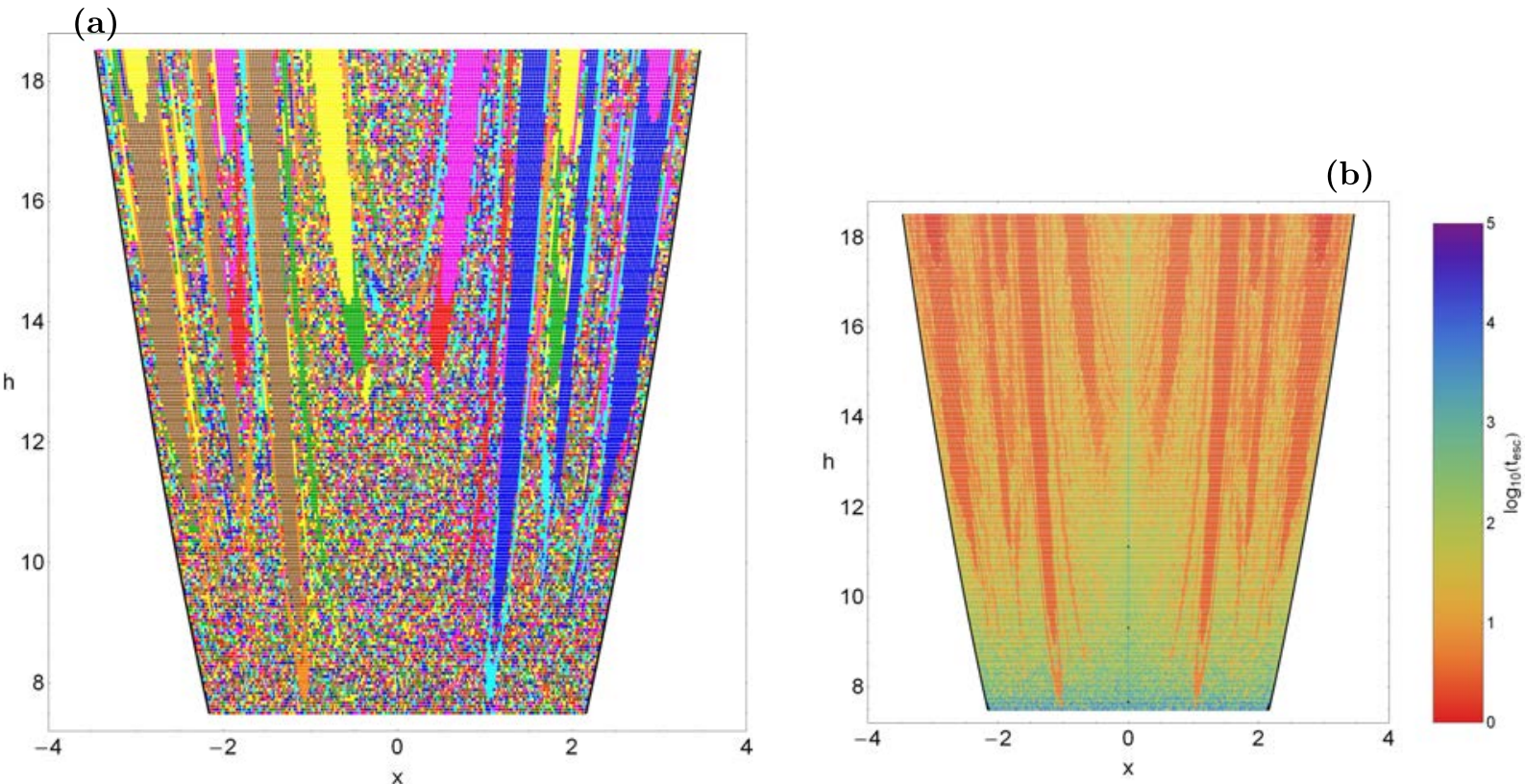}}
\caption{The structure of the $(x,h)$ plane when $z_0 = 1$ with respect to (a-left): escape channels and (b-right): the escape rates of orbits.}
\label{xh}
\end{figure*}

\begin{figure}
\includegraphics[width=\hsize]{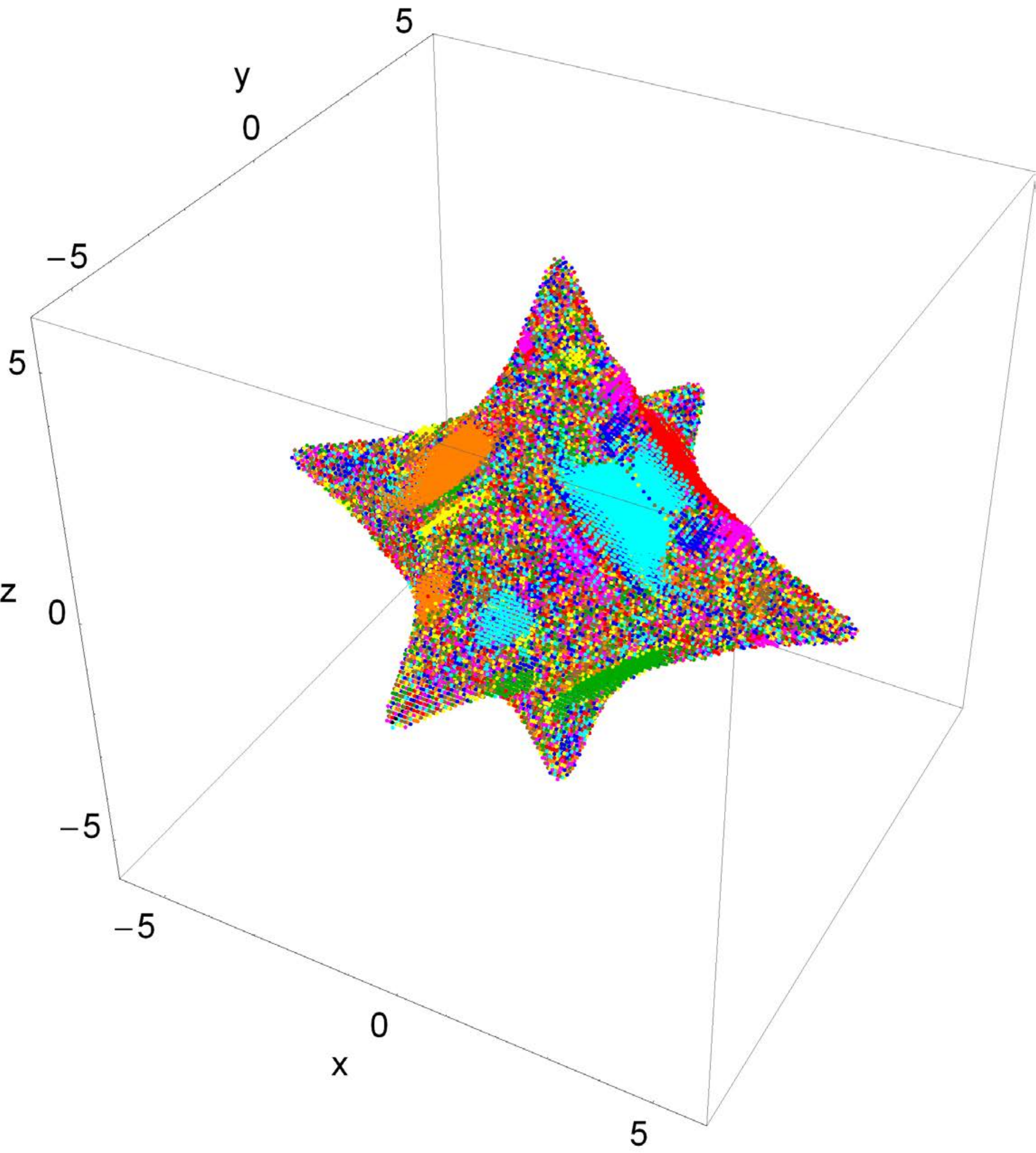}
\caption{A dense three-dimensional grid of about $10^5$ initial conditions $(x,y,z)$ inside the isopotential surface $V(x,y,z) = h$, when $h = 16.5$.}
\label{grid3d}
\end{figure}

Before closing this section, we would like to summarize and compare the results regarding the basins of escape. In Fig. \ref{basins} we show the evolution of the percentage of the total area on the $(x,\dot{x})$ plane corresponding to basins of escape, either broad regions or elongated spiral bands, as a function of the energy for the three initial values of the $z$ coordinate. It is seen, that for low values of $z_0$ (see Case I in subsection \ref{case1}, where $z_0 = 0.1$), only at high energy levels $(h > 17)$ we have weak indications of basins of escape corresponding to a tiny fraction of the grid, less than 5\%, while the vast majority of the grid displays a highly fractal structure. When $z_0 = 0.5$ on the other hand, basins of escape appear quite early for $h > 10$ and for larger values of the energy, they start to grow rapidly occupying about 65\% of the grid at the highest studied energy level. The growth of the extent of escape basins is considerably faster in the case where orbits are launched with high values of $z_0$. We observe, that for $z_0 = 1$ the rate of the basins of escape increases almost linearly obtaining high values very fast and covering about 90\% of the entire grid when $h = 18.5$. Therefore, one may conclude that for low values of $z_0$ the degree of fractalization is very high, while as we proceed to larger values of $z_0$ this behavior is restricted or confined and well-formed basins of escape take over the majority of the $(x,\dot{x})$ planes.

One may reasonably assume, that the basins of escape appear only at the $(x,\dot{x})$ plane and if we choose another subspace of the entire 6D space they might not be present. In order to give an answer to this, we decided to integrate orbits in three more planes and therefore check if the basins of escape are still present or not. Our results are presented in Fig. \ref{gridsm}(a-d) where the $(h = 16.5, z_0 = 1)$ model is examined. In particular, Fig. \ref{gridsm}a shows a grid of initial condition in the physical $(x,y)$ plane when $z = z_0$, $\dot{x_0} = \dot{z_0} = 0$, Fig. \ref{gridsm}b depicts the $(x,\dot{x})$ plane discussed previously in Fig. \ref{chansz10}, Fig. \ref{gridsm}c shows a grid of initial condition in the $(x,\dot{z})$ plane when $y = 0$, $z = z_0$, $\dot{x_0} = 0$, while in Fig. \ref{gridsm}d the grid of initial condition in the $(\dot{x},\dot{z})$ plane when $x = y = 0$ and $z = z_0$, while in all cases $\dot{y_0}$ is obtained from the energy integral (\ref{ham}). It is evident, that the escape basins are indeed present in all studied cases thus implying that their formation is an intrinsic property of the dynamical system regardless of the particular chosen two-dimensional plane of the 6D phase space.

Our numerical analysis reveals, that orbits with initial conditions $(x_0,\dot{x_0})$ inside the basins escape have either extremely low escape periods or they escape immediately from the system. At this point, we would like to emphasize this point a little bit. For this purpose, we constructed a three-dimensional plot shown in Fig. \ref{bas3d}, which combines results regarding both the escape channels and escape rates of the $(h = 16.5, z_0 = 1)$ model. It is seen, that obits belonging to escape basins possess low escape rates, while on the other hand, orbits inside the fractal region have relatively larger escape rates. With a much closer look we can see that orbits inside broad escape regions escape almost immediately, while orbits inside elongated bands spiralling around the center escape a little bit later.

\begin{figure*}
\centering
\resizebox{0.8\hsize}{!}{\rotatebox{90}{\includegraphics*{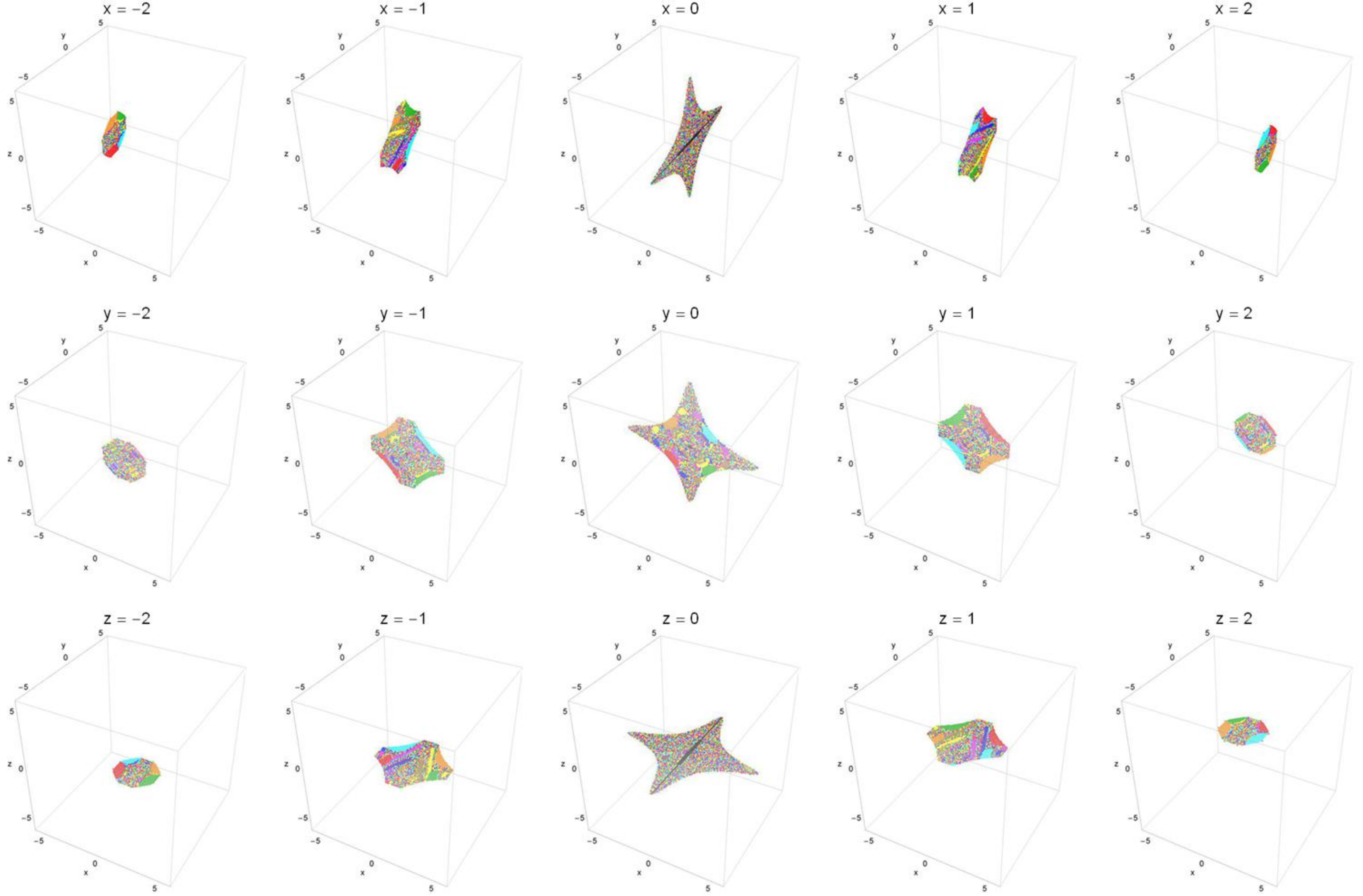}}}
\caption{A tomographic view of the solid grid of Fig. \ref{grid3d} showing slices on the primary $(x,y)$, $(x,z)$ and $(y,z)$ planes when $x,y,z = (-2,-1,0,1,2)$.}
\label{slices}
\end{figure*}

So far in our investigation, we chose discrete energy levels and we explored the escape process of orbits for three different values of the $z$-coordinate. It would be of particular interest, to study how the basins of escape evolve within a dense spectrum of values of the energy when $h = (h_{\rm esc}, 18.5]$. Therefore, we integrated orbits with initial conditions on the $x$ axis with $z_0 = 1$, $y_0 = \dot{x_0} = \dot{z_0}$, while the value of $\dot{y_0}$ is obtained from the energy integral. In Fig. \ref{xh}a we distinguish the initial conditions of orbits with respect to the escape channel, while Fig. \ref{xh}b shows the $(x,h)$ plane with the corresponding escape periods. The outermost black solid line denotes the Zero Velocity Curve (ZVC). We observe, that as we proceed to higher energy levels the presence of the escape basins becomes more and more prominent and at the same time the escape period reduces, while on the other hand the central region of the plane remains fractal all the way up. We would also like to point out, that both plots are qualitatively symmetrical with respect to the $x = 0$ axis.

Our last step will be to try to consolidate a general view of the escape process for different values of the $z$-coordinate. To obtain this, we defined a dense, three-dimensional grid of about $10^5$ initial conditions $(x_0,y_0,z_0)$ with $\dot{x_0} = \dot{z_0} = 0$ inside the isopotential surface when $h = 16.5$. As usual, the initial value of $\dot{y_0}$ was derived from the energy integral. In Fig. \ref{grid3d} we visualize the results of the numerical integration showing with the different color the escape channels. One may observe, once more, several basins of escape embedded in the surface of the grid. In this case, the grid is a three-dimensional solid therefore, only its surface is visible. However, we can penetrate inside the solid by using a tomographic-style approach. According to this method, we can again plot two-dimensional slices of the solid grid by defying specific levels to each primary plane. Fig. \ref{slices} shows the evolution of the structure of the primary $(x,y)$, $(x,z)$ and $(y,z)$ planes when $x,y,z = (-2,-1,0,1,2)$. We see, that the structure evolves rapidly and non uniformly (constant interplay between escape basins and fractal structure) thus implying that the escape process in this Hamiltonian system with three degrees of freedom (3 d.o.f) is, by all means, a very complex procedure.

\section{Discussion and conclusions}
\label{disc}

In the present paper, we tried to shed some light to the escape process in three-dimensional open Hamiltonian systems. We chose a simple potential of a perturbed harmonic oscillator which has a finite energy of escape. Beyond the escape energy the three-dimensional isopotential surfaces open and eight symmetric channels of escape appear. Our main objective was to reveal the nature of orbits and try to understand the escape process. In particular, we managed to distinguish between ordered/chaotic and trapped/escaping orbits and we also located the basins of escape leading to different escape channels, finding correlations with the corresponding escape times of the orbits. Our extensive and thorough numerical investigation strongly suggests, that overall escape process is very dependent on the value of the total orbital energy, as well as on the initial value of the $z$ coordinate.

Since a distribution function of the model was not available so as to use it for extracting the different samples of orbits, we had to follow an alternative path. We decided to restrict our exploration to a subspace (a 4D grid) of the whole 6D phase space. Specifically, we considered orbits with initial conditions $(x_0, z_0, \dot{x_0})$, $y_0 = \dot{z_0} = 0$, while the initial value of $\dot{y_0}$ is always obtained from the energy integral. Then, we defined a value of $z_0$, which was kept constant. Thus, we were able to construct again a 2D plot depicting the $(x, \dot{x})$ plane (we may call it a local phase plane) but with an additional value of $z_0$, since we deal with 3D motion. Following this approach, we were able to identify again regions of order and chaos. We defined for several values of the energy $h$, a dense grid of initial conditions $(x_0, \dot{x_0})$ regularly distributed in the area allowed by the value of the energy. In each grid the step separation of the initial conditions along the $x$ and $\dot{x}$ axis was controlled in such a way that always there are at least 25000 orbits. For the numerical integration of the orbits in each grid, we needed about between 2 minutes and 5.1 hours of CPU time on a Pentium Dual-Core 2.2 GHz PC, depending on the escape rates of orbits in each case.

In this work, we revealed the influence of the orbital energy $h$, as well as the initial value of the $z$ coordinate of three-dimensional orbits on the escape process. The main numerical results of our research can be summarized as follows:
\begin{enumerate}
 \item It was found, that more than 95\% of the computed orbits escape sooner or later, while only a tiny fraction of orbits corresponding mainly to stable periodic orbits remain trapped. For values of energy very close to the escape energy, the majority of escaping orbits are chaotic, while the percentage of chaotic escaping orbits reduces rapidly with increasing energy and for high energy levels the ordered escaping orbits is the most populated type of orbits. This behavior is more or less the same, regardless the particular initial value of the $z$ coordinate.
 \item A strong correlation between the extent of the basins of escape and $z_0$ was found to exists. Our numerical calculations indicate that for low values of $z_0$ the structure of the $(x,\dot{x})$ phase plane exhibits a large degree of fractalization and therefore orbits escape choosing randomly escape channels. As the value of $z_0$ increases however, the grid becomes less and less fractal and several basins of escape, either broad regions or elongated bands spiralling around the center emerge. The extent of these basins of escape is more prominent at high values of the energy, where they occupy about 90\% of the entire grid.
 \item It was also observed, that the escape times of orbits are directly linked to the basins of escape. Specifically, for low values of energy close to $h_{\rm esc}$, the escape periods are huge corresponding to tens of thousands of time units. This was justified, by taking into account that for these energy levels the width of the escape channels is very small, thus forcing the orbits to consume large time intervals inside the isopotential surface until they locate an exit and eventually escape to infinity. For higher energy levels on the other hand, several basins of escape are present so, the orbits escape very quickly within less than 10 time units, or even immediately from the system.
\end{enumerate}

The outcomes and the conclusions of the present research are considered, as an initial effort and also as a promising step in the task of understanding the escape mechanism of orbits in open three-dimensional Hamiltonian systems. Taking into account that our results are encouraging, it is in our future plans to expand our investigation in other more complicated potentials, focusing our interest in reveling the escape process of stars in galactic systems such as star clusters, or binary stellar systems.

\section*{Acknowledgments}

The author would like to thank the two anonymous referees for the careful reading of the manuscript and for all the apt suggestions and comments which allowed us to improve both the quality and the clarity of the paper.

\end{document}